\documentclass[]{aa}
\usepackage{graphicx,mytable,txfonts,natbib} 
\bibpunct{(}{)}{;}{a}{}{,} 

\begin{document}

\title{$L$- and $M$-band imaging observations of the Galactic Center region}


\author{T.~Viehmann\inst{1}
        \and A.~Eckart\inst{1}
        \and R.~Sch\"odel\inst{1}
        \and J.~Moultaka\inst{1}
        \and C.~Straubmeier\inst{1}
        \and J.-U.~Pott\inst{1,2}
}

\institute{I.~Physikalisches Institut, Universit\"at zu K\"oln,
           Z\"ulpicher~Str.~77, D-50937 K\"oln, Germany
           \and ESO, Karl-Schwarzschild-Str.~2, D-85748 Garching, Germany
}

\offprints{T. Viehmann, \email{viehmann@ph1.uni-koeln.de}}

\date{Received ; Accepted }

\abstract{
In this paper we present near-infrared $H$-, $K$-, $L$- and $M$-band photometry
of the Galactic Center from images obtained at the ESO VLT in May and August
2002, using the NAOS/CONICA ($H$ and $K$) and the ISAAC ($L$ and $M$)
instruments\thanks{based on observing run 69.B-0101(A) at the Very Large
Telescope (VLT) of the European Southern Observatory (ESO) on Paranal in
Chile.}.
The large field of view (70\arcsec$\times$70\arcsec) of the ISAAC instrument and
the large number of sources identified ($L-M$ data for a total of 541 sources)
allows us to investigate their colors, infrared excesses and the extended dust
emission.
Our new $L$-band magnitude calibration reveals an important offset to the
traditionally used ``standard'' calibrations, which we attribute to the use of
the variable star IRS~7 as a flux calibrator. Together with new results on the
extinction towards the Galactic Center
 \citep{scoville2003, raab2000},
our magnitude calibration results in  stellar color properties expected from
standard stars and removes any necessity to modify the $K$-band extinction. 
The large number of sources for which we have obtained $L-M$ colors allows us to
measure the $M$-band extinction to $A_M$=(0.056$\pm$0.006)$A_V$, i.e.
$A_M\approx A_L$, a considerably higher value than what has so far been assumed.
$L-M$ color data has not been investigated previously, due to lack of useful
$M$-band data. We find that this color is a useful diagnostic tool for the
preliminary identification of stellar types, since hot and cool stars show a
fairly clear $L-M$ color separation, with average $L-M$ colors of 0.46$\pm$0.17
for emission-line stars and $-$0.05$\pm$0.27 for cool red giants/AGB stars.
This is especially important if visual colors are not available, as in the
Galactic Center.\\
For one of the most prominent dust embedded sources, IRS~3, we find extended
$L$- and $M$-band continuum emission with a characteristic bow-shock shape. An
explanation for this unusual appearance is that IRS~3 consists of a massive,
hot, young mass-losing star surrounded by an optically thick, extended dust
shell, which is pushed northwest by wind from the direction of the IRS~16
cluster and Sgr~A*.
}

\titlerunning{$L$- and $M$-band Observations of the GC}
\authorrunning{Viehmann et al.}  

\maketitle


\section{Introduction}
\label{INTRO}

The Galactic Center stellar cluster shows some intriguing characteristics: it is
extremely dense, with an unusual observed stellar population consisting mainly
\citep[80\% of the $K\la$14 stars according to][]{ott1999}
of late-type red giants, many of which are suspected to lie on the asymptotic
giant branch (AGB), as well as young massive stars with energetic winds.
Spectroscopic measurements \citep[e.g.][]{krabbe1995, najarro1997} allow
identification of these two components, the late-type red giants and supergiants
showing strong 2.3~$\mu$m CO bandhead absorption and the massive, hot and windy
stars (``He stars'') exhibiting He and H emission lines.
The emission line stars appear to dominate energetically in the central few
arcseconds, where the bright IRS~16 cluster is located.
These stars are generally classified as Ofpe/WN9, although some of them might be
luminous blue variables (LBV) and a few show characteristics of Wolf-Rayet
stars.
A third, less numerous component of the Galactic Center stellar cluster consists
of luminous, extended objects with steep, red and featureless ($K$-band) spectra
and a strong infrared excess.
These sources were considered as potential YSO candidates
\citep{clenet2001,ott1999,krabbe1995}.
Recently, however, they have been successfully described as bow-shock sources
\citep{tanner2002, tanner2003, rigaut2003, eckart2004}, with the implication
that the stars powering these luminous infrared sources are windy stars of the
types mentioned above.

We present multi-band ($H, K, L$ and $M$) photometry of the Galactic Center,
obtained with the ESO VLT, in order to further investigate the properties of the
Galactic Center stellar cluster.
The large number of sources available for color analysis allows for improved
statistics, resulting in a new $L$-band calibration which eliminates anomalous
color effects found in other surveys of the Galactic Center stellar cluster
\citep[e.g.][]{clenet2001, sb1996}.
The large field of view (70\arcsec$\times$70\arcsec) of the ISAAC instrument
allows us to compare the immediate Galactic Center cluster population with the
stellar population further out, with separations from SgrA* of up to about half
an arcminute.
This comparison results in a confirmation of the new extinction values of 
\citet{scoville2003}.
Previous (multi-band) imaging studies of the Galactic Center, besides being
limited in their field of view to e.g. 13\arcsec$\times$13\arcsec
\citep{clenet2001, davidge1997}, 
appear to have neglected the $M$-band (4.66~$\mu$m), which shows some
interesting properties, which we discuss in this paper.

\section{Observations and data reduction}
\label{obs}

 \subsection{ISAAC observations}
On May 23rd -- 30th 2002, $L$- and $M$-band (3.78~$\mu$m and 4.66~$\mu$m)
imaging observations of the Galactic Center were obtained with the ISAAC
infrared imager at the ESO VLT (Unit Telescope 1, Antu), as part of a joint
IR/X-ray monitoring program of Sgr~A* \citep{eckart2003sgra, baganoff2003}.
The seeing-limited images were taken in ISAAC's higher resolution mode, with a
pixel scale of 0.071\arcsec and a field of view of
72.6\arcsec$\times$72.6\arcsec \citep[see][]{isaacmanual}.
Seeing conditions varied during the campaign, resulting in angular resolutions
between 0.4\arcsec and 1.2\arcsec.
The observations were made in batches of 25 to 40 pairs of frames (with chopping
and nodding), with an integration time of 0.99~s per frame.

 \subsection{NAOS/CONICA observations}
During the science verification phase of the commissioning and observatory
preparation of the NAOS/CONICA adaptive optics assisted imager/spectrometer at
the ESO VLT (UT4, Yepun), $H$- and $K_s$-band (1.6~$\mu$m and 2.1~$\mu$m) images
of the Galactic Center were obtained on 30 August 2002, with a FWHM of 43 and 56
mas, respectively, and a pixel scale of 0.0132\arcsec/pixel
\citep{schoedel2002nature, schoedel2003}.
The total integration times were 1500~s and 1200~s, respectively.
Since the $K-K_s$ color difference 
is an order of magnitude smaller than the photometric accuracy that we reach, no
further distinction between $K$ (2.2~$\mu$m) and $K_s$ will be made in this
paper.
Since the purpose of the NAOS/CONICA images in this context was to gain
additional color information for analysing the seeing-limited ISAAC $L$- and
$M$-band images, the pixel scale of the NAOS/CONICA images was reduced to the
corresponding 0.071\arcsec/pixel via re-binning.
Further data reduction was then performed using the same methods as with the
ISAAC images.

 \subsection{Data reduction}
\begin{figure*}
\centering
 \includegraphics[width=17cm]{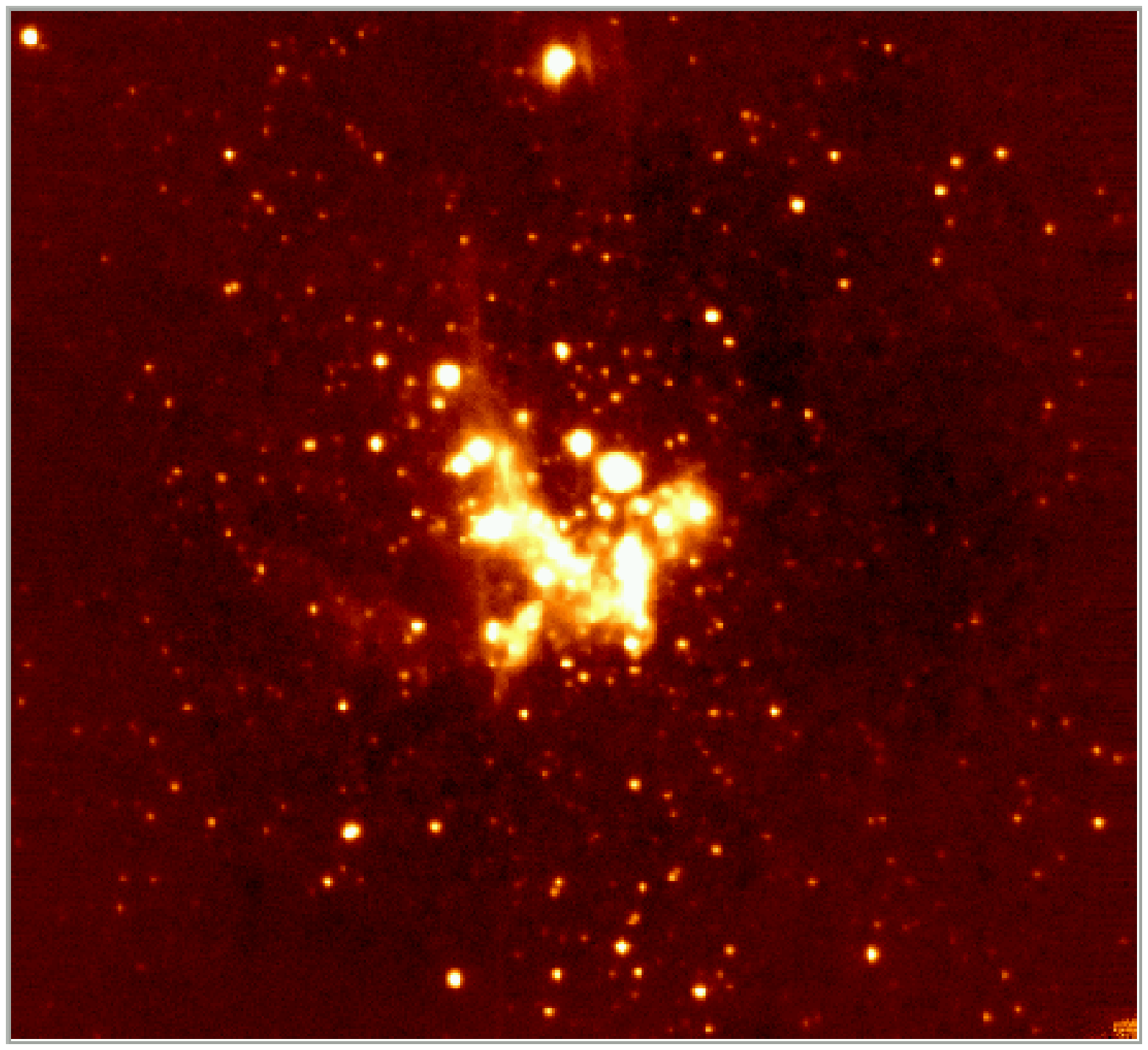}
 \caption{ISAAC $M$-band final image of the Galactic Center. The field of view
 is 72.6\arcsec$\times$72.6\arcsec.}
 \label{fullmband}
\end{figure*}

All image processing was performed using routines from the IRAF software
package.
The data reduction procedures employed were standard:
After flat field correction, the paired images were subtracted from each other
and subsequently shifted to a common reference frame.
All such subtracted images from each batch were then combined by calculating the
median, which also effectively removed cosmic rays and bad pixels.
The final images were then created by coaddition of the median images with the
best effective seeing, i.e. better than 0.6\arcsec, resulting in total
integration times of 229.7~s for the $L$-band image and 162.4~s for the $M$-band
image.
The final $M$-band image is shown in Fig.~\ref{fullmband}.
Finally, PSF-fitting photometry was performed on all four final images using the
IRAF/DAOPHOT software package.

 \subsection{Photometry}
 \label{calib}

The observing conditions for both the ISAAC and the NAOS/CONICA
observations were generally photometric and are adequate for relative
photometry.
For the $L$- and $M$-band case, we linked the calibration of the IRAF/DAOPHOT
PSF-fitting results to observed flux density reference stars.
For the $H$- and $K$-band, we relied on data taken from the literature
\citep{clenet2001, depoy1990}.

  \paragraph{$L$- and $M$-Band Magnitude Calibration\\}
For calibration purposes, the standard stars HD~130163 \citep[$m_L=$6.856,
$m_L-m_M\approx0$,][]{vdbliek1996cat} and HR~5249
\citep[$m_M=$4.53,][]{vdbliek1996cat} were imaged together with science
observations on May 28th 2002, using identical techniques.
The airmass difference was small (maximum 0.2), resulting in corrections smaller
than 0.04 magnitudes, which are insignificant given the uncertainties in the
photometry (see section \ref{lotsofnumbers}).
These observations were used to calibrate the final $L$- and $M$-band images,
using aperture photometry on approximately 30 isolated stars.

  \paragraph{$K$-Band Magnitude Calibration\\}
Reference $K$-band magnitudes for some of the bright IRS sources were taken from
\citet{clenet2001}, who fixed their zero-point with the photometry of
\citet{ott1999}.
The differences between the DAOPHOT magnitudes and the values of
\citet{clenet2001} were averaged, resulting in a calibration value
which was subtracted from the DAOPHOT magnitudes.
Saturated sources were excluded from the determination of the calibration
magnitudes.

  \paragraph{$H$-Band Magnitude Calibration\\}
The main problem encountered in calibrating the $H$-band data was saturation.
All the bright $H$-band sources (for which magnitudes are available) are
saturated in the NAOS/CONICA image, whereas no magnitudes for the weaker
unsaturated sources could be found in the literature.\\
The solution employed here was secondary calibration using an $H$-band mosaic
made from images taken with the SHARP speckle imager in 1992 \citep{eckart1992}.
This image was flux calibrated using data for the brighter IRS sources
\citep{depoy1990}.
Subsequently, the NAOS/CONICA image was calibrated with the flux densities from
weaker, isolated (i.e. clearly separated from their neighbors) sources in the
SHARP mosaic image.\\

The total uncertainty of the photometry results is estimated as 0.25
magnitudes in the $H$- and $K$-band and 0.15 magnitudes in the $L$- and $M$-band.
The difference is due to the different calibration procedures and the different
shape of the respective PSFs of the ISAAC and NAOS/CONICA images, resulting in
less accurate fit results for the latter.
The limiting magnitudes of this analysis are 12.73 ($M$-band), 14.47 ($L$-band),
17.66 ($K$-band) and 20.34 ($H$-band).

 \subsection{Additional data}
The field of view of the NAOS/CONICA images (approximately
17.8\arcsec$\times$17.8\arcsec) is much smaller than the ISAAC field of view.
In order to check the properties of sources outside this region, additional $H$-
and $K$-band images were analysed.

  \paragraph{ISAAC $K$-band image\\}
During the 1999 ISAAC spectroscopy campaign \citep{eckart1999},
a $K$-band image of the Galactic Center was obtained, using ISAAC's lower
resolution mode \citep[pixel scale 0.148\arcsec, field of view approximately
150\arcsec$\times$150\arcsec, see][]{isaacmanual}.

  \paragraph{Gemini data\\}
For $H$-band, we also used publicly released Gemini data from the 2000 Galactic
Center Demonstration Science Data Set \citep[using the Quirc camera and the Hokupa'a
adaptive optics system, see][]{coterarigaut2001}.

In order to limit the amount of work required, photometry on these datasets was
performed only for the brighter sources in the $L$-band image ($m_L<$9.5) and an
exemplary selection of approximately 15 fainter sources in order to verify that
the faint stars do not show systematically different color properties than the
bright stars.
The images were calibrated relative to our NAOS/CONICA data.

\section{Results}
\label{results}

 \subsection{Photometry}
\label{lotsofnumbers}
  Table \ref{fat} summarizes the photometry of those sources within the
  NAOS/CONICA field of view (17.8\arcsec$\times$17.8\arcsec) that are detectable
  in at least one of the ISAAC images.
  Table \ref{outside} summarizes the photometry of the sources outside the
  NAOS/CONICA field of view, for which therefore only $L$- and $M$-band data are
  available.\\

\begin{figure*}
\centering
 \includegraphics[width=17cm]{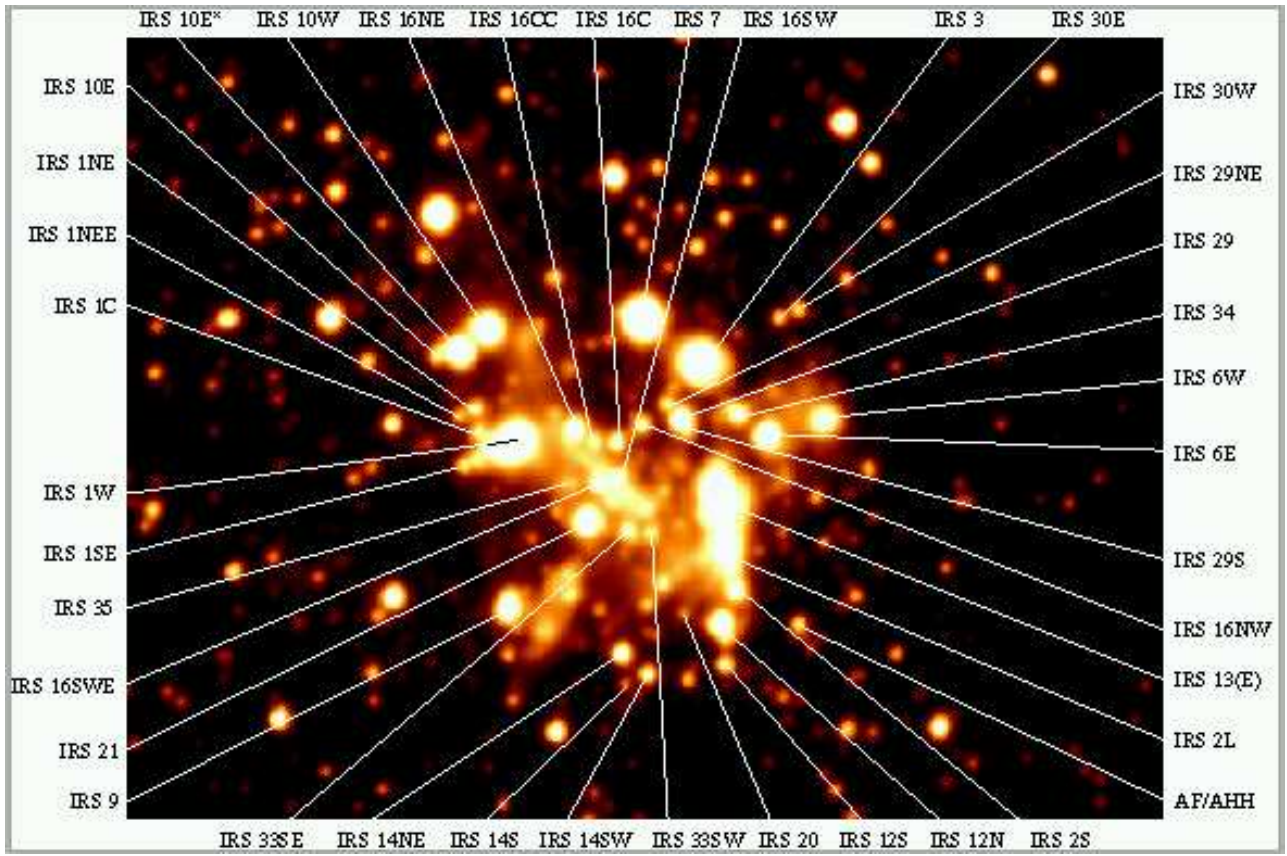}
 \caption{Part of the ISAAC $L$-band final image of the Galactic Center showing
 most of the named sources.}
 \label{srcinl}
\end{figure*}

  Notes:
  \begin{itemize}
   \item The named sources are identified in Fig.~\ref{srcinl}.
   \item The position offset is given in arcseconds relative to IRS~16NE. 
    In order to calculate the position offsets with respect to the location 
    of the infrared/radio source Sgr~A* \citep{reid2002, schoedel2003}, one has
    to add 2.83\arcsec in right ascension and $-$0.91\arcsec in declination.
    Given the larger field of view, the positions were derived from our 
    best seeing L-band map and aligned with respect to the radio positional 
    reference frame for the Galactic Center as given by \citet{reid2002}.
   \item For the astrometry, the pixel scale from the ISAAC user manual was used
    \citep[see][]{isaacmanual} with minor corrections based on the comparison of
    our positions to those of \citet{ott1999}.
    From this comparison, we estimate the absolute position error as
    approximately 0.09 arcseconds.
    Relative positions in our images can be derived with a much higher precision
    of at least half a pixel, i.e. $\la$0.035\arcsec.
   \item If a source was not detected at a given wavelength, the maximum
    detected magnitude in that band is given as a lower limit (i.e. upper
    brightness limit).
    These cases were excluded from the corresponding color-color diagrams.
  \end{itemize}

 \subsection{New $L$-band calibration}
\label{newl} 
  Since our $L$-band calibration introduces an offset of approximately 0.5~mag
  to the frequently used calibration of \citet{blum1996}, some comments are
  necessary.
  
  Calibrating the $L$-band image with reference magnitudes taken from
  \citet{clenet2001}, as for the NAOS/CONICA $K$-band image (see section
  \ref{calib}), resulted in $(K-L)_{\mathrm{obs}}$ colors which were \emph{at
  least} 0.3 magnitudes too blue.
  An examination of the color-color-diagrams showed this effect to be
  systematic, since \emph{the whole population} was blueshifted in $K-L$,
  whereas the $H-K$ colors lie within the expected range for stellar sources.
  It therefore seems unlikely that the $K$-band calibration is wrong, especially
  considering that it is based on stars shown to be non-variable \citep{ott1999,
  clenet2001}.
  Directly calibrating the $L$-band image with the observed calibration star
  instead of adjusting the zero-point with already published photometry does not
  show this effect.
  We note that \citet{clenet2001} also found unusually blue $K-L$ values,
  which they attributed to a lower $K$-band extinction. If the extinction law
  remains unchanged, this implicitly suggests a lower extinction $A_V$, which
  our observations can confirm (see section \ref{colext}).
  
  The $L$-band data and thus the 
  calibration given by \citet{blum1996} are taken from \citet{depoysharp1991}.
  Their calibration (which is not discussed in detail in their 1991 publication)
  was fixed with the star IRS~7, for which \citet{blum1996, tamura1996} and
  \citet{ott1999} have shown that it is variable with an amplitude of of
  approximately 1~mag in $K$-band.
  It seems that the offset to the other frequently applied calibration by
  \citet{sb1996} is also due to their use of the variable source IRS~7 as flux
  calibrator.
  Our new $L$-band calibration, however, provides a clear improvement to this
  point, especially since it does not rely on any (non-)variability assumptions
  for the observed sources and does not result in unexpected $K-L$ color
  effects.
  We finally note that a higher extinction than the assumed $A_V$=25~mag would
  result in even bluer $K-L$ colors, requiring an even larger offset to the
  $L$-band calibrations mentioned above.
  
 \subsection{Colors and extinction}
\label{colext}

\begin{figure*}
   \begin{center}
    \includegraphics[width=17cm]{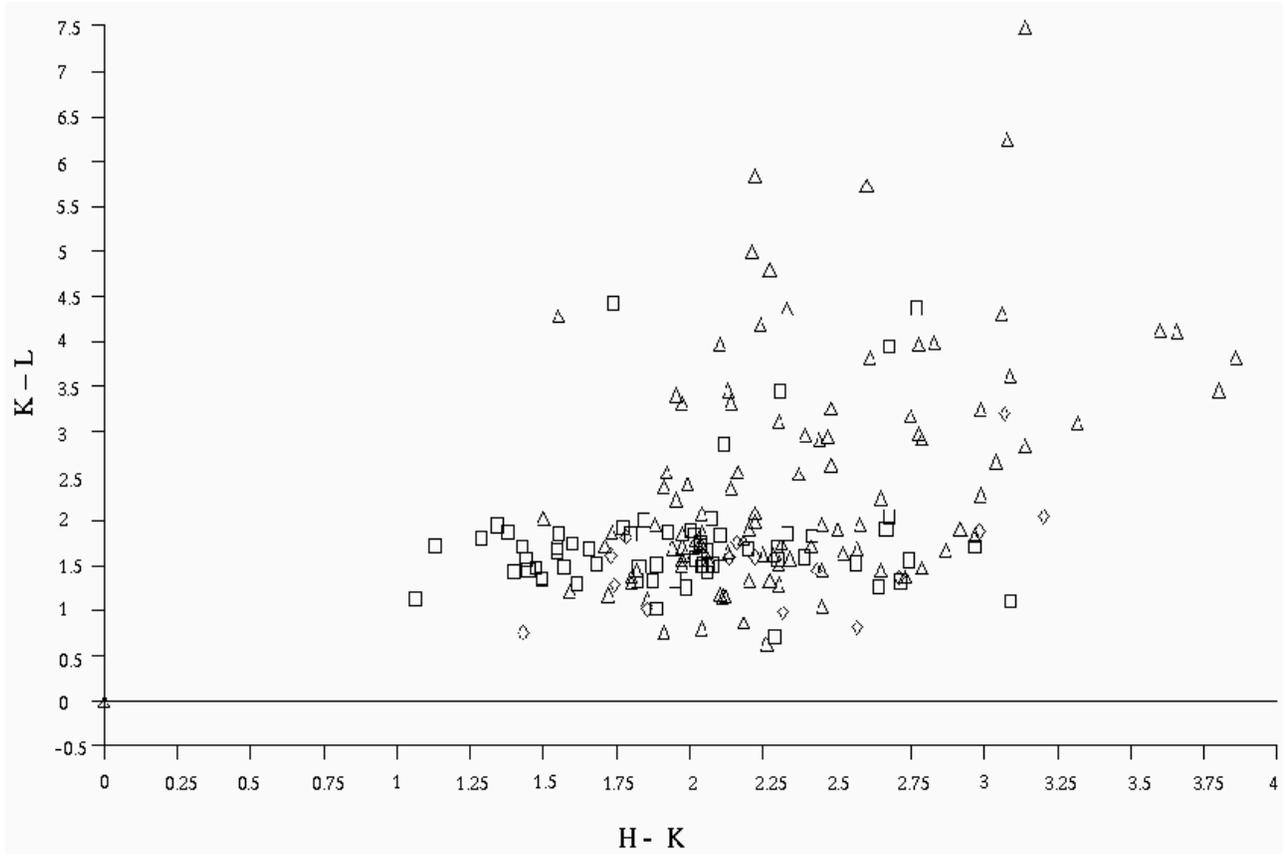}
    \caption{Two-color diagram showing $(K-L)_{\mathrm{obs}}$ versus
    $(H-K)_{\mathrm{obs}}$. The triangles represent sources from the inner
    18\arcsec$\times$18\arcsec, the squares represent brighter sources
    ($m_L<$9.5) and the diamonds a selection of fainter sources from outside
    this inner region.}
\label{hkl}
   \end{center}
\end{figure*}

  For de-reddening purposes, the extinction law of \citet{rl1985} was adopted,
  i.e. $A_H=0.175A_V$, $A_K=0.112A_V$, $A_L=0.058A_V$ and (initially)
  $A_M=0.023A_V$.
  Since recent work by \citet{scoville2003} indicates that the extinction toward
  the Galactic Center exhibits a broad minimum over the area of the IRS~16
  Cluster and the Northern Arm, their value of $A_V=$25~mag was assumed, instead
  of the traditional average value of $A_V=$27 or even 30~mag \citep{rieke1987}.
  This choice is supported by the colors of the stars observed, especially $K-L$
  (see section \ref{newl}), since they would otherwise be too blue.\\
  In the $HKL$ color-color diagram (Fig.~\ref{hkl}), the majority of sources
  form a cluster in the area populated by stars
  \citep[for intrinsic colors see][]{koornneef1983}, with a fairly sharp
  boundary towards bluer colors.
  This is satisfactory, since the majority of sources exhibit colors that lie in
  the possible color range for stellar sources. Only \emph{very few} objects
  show bluer colors, which can be seen as exceptions.
  Quite a few objects appear to be strongly reddened, especially within the
  inner area covered by the NAOS/CONICA images.
  This is not surprising, since the dust and gas of the `mini-spiral' passes
  through this region.
  
  The conclusion to be drawn here is that the newer findings concerning
  extinction \citep{scoville2003}, together with our new $L$-band calibration,
  provide a consistent picture without requiring modifications of the 
  extinction law at this point.

  \subsubsection{Comparison of outer and inner field}
\begin{figure*}
   \begin{center}
    \includegraphics[width=17cm]{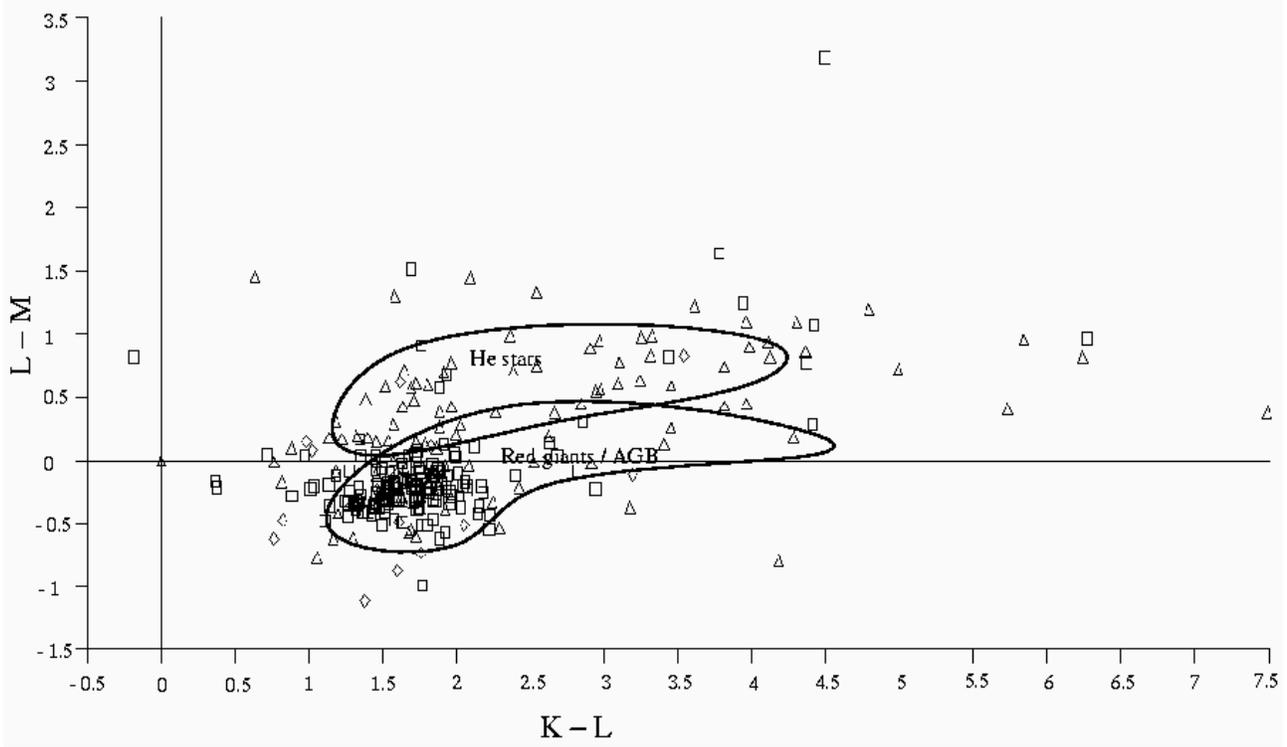}
    \caption{Two-color diagram showing $(L-M)_{\mathrm{obs}}$ versus
    $(K-L)_{\mathrm{obs}}$. The triangles represent sources from the inner
    18\arcsec$\times$18\arcsec, the squares represent brighter sources
    ($m_L<$9.5) and the diamonds a selection of fainter sources from outside
    this inner region. The marked areas fully contain the spectroscopically
    identified He stars and red giants / AGB stars from Table \ref{fat}.}
    \label{klm}
   \end{center}
\end{figure*}

   With the exception of some highly reddened sources (e.g. IRS~1W, IRS~3,
   IRS~10, IRS~21) found almost exclusively within the inner field (i.e. the
   18\arcsec$\times$18\arcsec of the NAOS/CONICA field of view), the stars from
   the inner and outer regions cluster in the same area of the two-color
   diagrams, with the clustering much more pronounced for the stars from the
   outer region.
   The highly reddened sources from the inner field are apparently
   obscured/reddened by gas and dust from the mini-spiral (either directly or by
   a bow-shock type interaction).\\
   The $H-K$ and $K-L$ colors obtained from our data, especially if one excludes
   unusual stars such as Wolf-Rayet stars etc., confirm the results of
   \citet{scoville2003} for the lower extinction (we assumed $A_V$=25~mag)
   towards the IRS~16 cluster and the Northern Arm of the mini-spiral, since the
   colors -- after dereddening -- agree with the intrinsic colors of stars given
   by \citet{koornneef1983}.
   Any higher extinction would lead to significantly bluer intrinsic colors,
   which cannot be explained in a straightforward manner.
   Our data indicates that this relatively low extinction applies to the
   complete region covered by the (high-resolution) ISAAC images, i.e. the inner
   70\arcsec$\times$70\arcsec.
   This can be inferred from the two-color diagrams (Figs.~\ref{hkl} and
   \ref{klm}), since the populations of the inner and outer region cluster in
   the same area of these diagrams.
   If the extinction were to increase significantly within the ISAAC high
   resolution field of view, the `outer' population (or at least a significant
   part of it) should show systematically redder colors than the main part of
   the `inner' population, which is clearly not the case.
   The higher extinctions of 35 to 50 mag reported by \citet{scoville2003} thus
   apply to areas outside the inner 70 arcseconds, which is consistent with the
   location of the circum-nuclear molecular ring, which has a sharp inner edge
   \citep{latvakoski1999}.

  \subsubsection{Extinction in $M$-band}
   Our $M$-band calibration, using two calibration stars of significantly
   different magnitude (see section \ref{calib}), provides rather surprising
   results:
   the stars appear too blue by about 0.7~mag in $L-M$, if one applies the
   reddening law of \citet{rl1985}, i.e. $A_M$=0.023$A_V$.
   However, the ISO/SWS Galactic Center extinction measurements of
   \citet{lutz1996} and \citet{lutz1999}, while generally in agreement with the
   extinction law of \citet{rl1985}, permit considerably higher (or lower)
   extinctions in the 4-5~$\mu$m regime within their uncertainties.
   This is especially true if one considers that their extinction values
   increase again longward of 5~$\mu$m \citep[see also][]{raab2000}.
   Visual inspection of the distribution of $L-M$ colors in the color-color
   diagram (Fig.~\ref{klm}) suggests that $A_M\approx A_L$, since the bulk of
   the $L-M$ colors observed then agree with the expected $L-M$ colors for
   stellar sources \citep{koornneef1983}.

\begin{figure}
 \resizebox{\hsize}{!}{\includegraphics{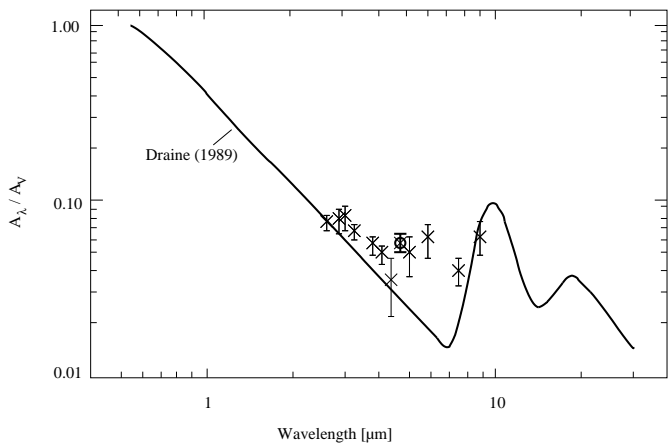}}
 \caption{Extinction towards Sgr~A* according to \protect\citet{lutz1996}. The
 data points are derived from SWS H recombination line data, while the solid
 curve is the extinction law of \protect\citet{draine1989} for standard
 graphite-silicate mixes. The marked data point is our new $M$-band result.}
 \label{lutzi}
\end{figure}

   Our extensive $L-M$ color data available for a total of 541 sources allows us
   to quantify the $M$-band extinction more precisely, assuming that the
   theoretical intrinsic colors given by \citet{koornneef1983} are valid.
   From the mean observed and theoretical intrinsic $L-M$ colors averaged over
   all types of stars we can calculate the $M$-band extinction via:
\[\frac{A_M}{A_V}=\frac{1}{A_V}(\langle L-M\rangle_{\mathrm{Koornneef}}-\langle L-M\rangle_{\mathrm{obs}}+A_L)=0.056\pm0.006.\]
   Here we quote a 3$\sigma$ uncertainty.
   This result is plotted in Fig.~\ref{lutzi} and agrees very well with the
   above estimate of $A_M\approx A_L$, with $A_L$=0.058$A_V$ following
   \citet{rl1985}.
   The possible luminosity-induced bias (observed stars are bright) is $<0.003$
   and therefore not significant given the uncertainties.

   Assuming for simplicity a Gaussian distribution of the $L-M$ values for
   simplicity, we find a corresponding standard deviation of
   $\sigma_{L-M}$=0.51, which allows the calculation of the contribution
   $\sigma_{\mathrm{ext}}$ of screening and source intrinsic effects to the
   $L-M$ distribution.
   The uncertainties of the distribution are given by
\[\sigma_{L-M}^2=\sigma_L^2+\sigma_M^2+\sigma_{\mathrm{Koornneef}}^2+\sigma_{\mathrm{ext}}^2\
,\]
   where $\sigma_{\mathrm{Koornneef}}$=0.16 represents the scattering of
   theoretical intrinsic colors over the whole range of stellar types as listed
   by \citet{koornneef1983}.
   Here $\sigma_L$ and $\sigma_M$ are the uncertainties in the L-and M-band
   magnitudes.
   We find $\sigma_{\mathrm{ext}}=$0.44, which implies that the \emph{main}
   contribution to the width of the $L-M$ color distribution is due to screening
   effects and source-intrinsic reddening (i.e. local dust concentrations etc.).
   The main caveat here is that the $L-M$ color distribution is not Gaussian
   (there are two principal stellar types with different intrinsic colors, i.e.
   red giants and young, massive emission-line stars). However, the width of the
   distribution due to different stellar type components is considerably smaller
   than the scattering observed.
   Nevertheless, this analysis is useful as a preliminary estimate, and further
   $M$-band studies of the Galactic Center are necessary to clarify the
   situation.

  \subsubsection{Colors of He-stars and red giants}
   A comparison of the $L-M$ colors of the named sources of known type (i.e.
   He-stars or cool red giants/AGB) shows a fairly clear color separation, with
   the cool stars significantly \emph{bluer} (see Table \ref{fat} and
   Fig.~\ref{klm}).
   This result is in general agreement with the colors given by
   \citet{koornneef1983}:
   Hot stars have on average an $(L-M)_{\mathrm{intrinsic}}$ of $\approx$-0.05,
   while cool giants have $(L-M)_{\mathrm{intrinsic}}$ values of typically -0.2
   to -0.3.
   In our sample, the average observed $(L-M)_{\mathrm{obs}}$ values are
   0.46$\pm$0.17 for the emission-line stars and $-$0.05$\pm$0.27 for the cool
   stars (ten of each, see Table \ref{fat}), with the standard deviation as
   uncertainty.
   Therefore, the $L-M$ color, and more particularly the $H-K$ versus $L-M$
   color-color diagram is a reasonably good diagnostic tool for preliminary
   identification of these stellar types.
   Applying this $L-M$ criterion to our sample, we find that the emission-line
   stars show a clear concentration in the inner region, with only about ten
   emission-line-star candidates in the
   area outside the 18\arcsec$\times$18\arcsec NAOS/CONICA field of view
   (containing 166 objects with full photometry data), as opposed to at least
   forty such objects in the inner region (197 objects in total).
      
   $K-L$ colors are much less useful in this context, since the $K-L$ color
   separation is comparatively low.
   Naturally, spectroscopic evidence is required for a definite classification.
   
   An additional comment can be made on the dereddened (i.e.\ intrinsic)
   $(K-L)_{\mathrm{intrinsic}}$ and $(L-M)_{\mathrm{intrinsic}}$ colors of the
   He-stars.
   These appear to be slightly but systematically redder ($\approx$0.2~mag) than
   expected for hot stars, whose $(K-L)_{\mathrm{intrinsic}}$
   and $(L-M)_{\mathrm{intrinsic}}$ colors are approximately 0 to -0.1.
   This indicates the presence of an infrared excess.
   A possible explanation for such an excess could be that the He-stars all show
   signs of an interaction with the dusty Galactic Center ISM.
   This excess could be of a similar nature but much less pronounced than the
   bow shock sources along the mini-spiral \citep{tanner2002, tanner2003,
   rigaut2003, eckart2004, clenet2004}.
   Our result is in general agreement with the findings of \citet{davidge1997},
   who indicate that the $K-L$ colors of the He-stars are clearly redder than
   those of known emission line stars in the LMC.

 \subsection{IRS~3}
\label{irs3}
\begin{figure*}
\centering
 \includegraphics[width=17cm]{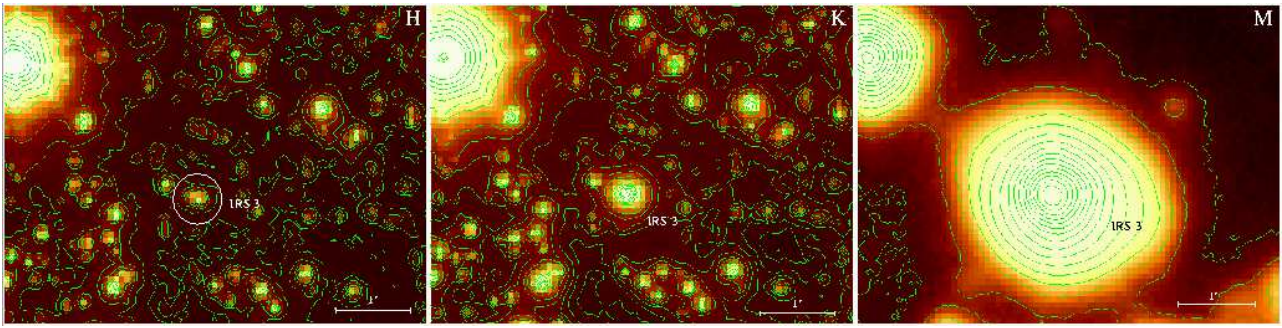}
 \caption{$H$-,$K$- and $M$-band view of IRS~3 (encircled in the $H$-band
 view). The bright source to the northeast (upper left) is the supergiant IRS~7.
 A histogram equalization color scale is used, while the contour lines follow a
 logarithmic scale in order to show the IRS~3 source structure as clearly as
 possible.}
 \label{irs3hkm}
\end{figure*}

\begin{figure}
 \resizebox{\hsize}{!}{\includegraphics{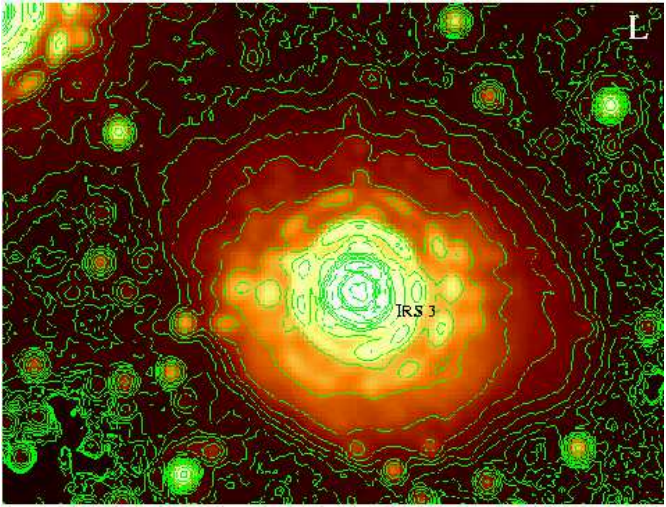}}
 \caption{$L$-band view of IRS~3 with NAOS/CONICA \protect\citep{eckart2004}.
 The circular structures surrounding the bright sources are due to the adaptive
 optics. Note the asymmetry in the lower contour lines producing a
 bow-shock-like effect at the southern edge of the extended emission.}
 \label{irs3l}
\end{figure}

  One of the most intriguing mid-infrared sources in the Galactic Center region
  is IRS~3.
  In the $L$- and $M$-band, this is the most extended source observed here, and
  also one of the brightest (m$_M=$3.4).
  It is much fainter in the $K$-band, hardly visible in the $H$-band
  (m$_K=$10.6, m$_H$=15.0), and also hardly extended at these shorter
  wavelengths (see Figs.~\ref{irs3hkm} and \ref{irs3l}).
  Consequently, IRS~3 appears to be dominated by dust emission.
  According to \citet{krabbe1995}, IRS~3 has an almost featureless, red
  continuum $K$-band spectrum.
  Also IRS~3 emits strongly at 10~$\mu$m \citep[e.g.][]{gezari1985}.\\
  In our analysis, the $M$-band photocenter of IRS~3 is shifted to the northwest
  by $\sim$0.16\arcsec with respect to the $H$-,$K$- and $L$-band positions.
  This angular separation is well above the relative positional uncertainty of
  0.035\arcsec we estimated above and thus not a mere artifact.
  A visual inspection of the maps clearly shows, however, that the two sources
  detected by the PSF-fitting algorithm are not distinct sources, but a
  consequence of the extended flux density distribution.
  This shift of the photocenter of the extended $M$-band emission relative to
  that at shorter wavelengths, combined with the clearly visible asymmetry of
  the lower contour lines in the NAOS/CONICA $L$-band image (Fig.~\ref{irs3l})
  results in a bow-shock-like appearance, similar to the sources discussed by
  \citet{tanner2003}, although less distinct than the remarkable source
  mentioned by \citet{clenet2004}.
  The asymmetry is the consequence of a compression of  the extended emission at
  its southern edge, resulting in a steeper flux density gradient (``bunched''
  contour lines).
  A possible explanation for this unusual appearance is that IRS~3 consists of a hot
  mass-losing star, e.g. a young, dust-embedded O star \citep{krabbe1995}
  or a dusty protostar or Wolf-Rayet star \citep{tanner2003, horrobin2004},
  surrounded by a very thick, extended dust shell, which is pushed northwest by
  wind from the direction of the IRS~16 cluster and Sgr~A*.
  \cite{gezari1985} give a temperature of $\approx$400~K for this strong
  $L$-band source.
  The $H$- and $K$-band emission is dominated by the stellar source and
  therefore not extended.
  The dust shell must be very thick, however, since IRS~3 is a faint source in
  $K$ and especially $H$.
  The shift in the $M$-band emission is due to part of the dust shell being
  extended to the northwest.
  It should be possible to clarify the nature of IRS~3 through $L$- and
  $M$-band (and possibly longer wavelength) studies at higher resolution, i.e.
  using adaptive optics or interferometric methods that are becoming available
  in the infrared \citep[e.g. VLTI or the upcoming LBT, see][]{eckart2003vlti}.

\section{Conclusion}
\label{conc}

We have presented multi-band infrared photometry of the Galactic Center,
covering the large field of view (70\arcsec$\times$70\arcsec) of the ISAAC
instrument at the VLT.
Our independent $L$-band magnitude calibration, relying on the robust method of
observing standard calibration stars, reveals an offset to the traditionally
used ``standard'' calibrations \citep{depoysharp1991, blum1996, sb1996}.
We attribute this offset to the use of the variable star IRS~7 \citep{blum1996,
tamura1996, ott1999} as a flux calibrator.
This offset is important when investigating colors.

Together with new results on the extinction towards the Galactic Center
\citep{scoville2003}, our magnitude calibration makes the stars redder in $K-L$,
which results in normal stellar properties.
There is no longer any necessity to modify the $K$-band extinction in order to
explain strange blue colors.

Color analysis of our large sample of sources is consistent with the results
obtained by \citet{scoville2003} concerning the extinction towards the Galactic
Center, since we can compare the color distributions of sources lying within an
inner 18\arcsec$\times$18\arcsec region with sources located further out, up to
the 70\arcsec$\times$70\arcsec limit of the field of view of the ISAAC camera.
This comparison reveals no systematic effects, indicating that the average
extinction remains at the lower value of $A_V=$25~mag up to the sharp inner edge
of the circum-nuclear molecular ring.

Our $M$-band data indicates that the extinction in the $M$-band is higher than
predicted by the extinction law of \citet{rl1985}.
The large number of sources for which we have obtained $L-M$ colors allows a
measurement of the $M$-band extinction, assuming that the intrinsic colors given
by \citet{koornneef1983} are valid.
This measurement gives $A_M$=0.056$A_V$, i.e. $A_M\approx A_L$, a considerably
higher value compared to what has been assumed so far.

Hot and cool stars show a fairly clear $L-M$ color separation
\citep{koornneef1983}, with the hotter stars exhibiting \emph{redder} colors.
Our results show that the Galactic Center sources follow this behavior,
for which the $H-K$ versus $L-M$ color-color diagram (if available) is a useful
diagnostic tool for preliminary identification of stellar types.

Several sources within the central few arcseconds are dust embedded and show
extended $L$- and $M$-band emission.
For one of the most prominent dust-embedded sources, IRS~3, we find extended
$L$- and $M$-band continuum emission with a characteristic bow-shock-like
appearance, including a shifted $M$-band photocenter relative to the other
wavelengths.
A possible explanation for this unusual appearance is that IRS~3 consists of a
massive, hot, young and mass-losing star surrounded by an optically thick,
extended dust shell, which interacts with the strong wind from the direction of
the IRS~16 cluster and Sgr~A*.
As a result of this the dust shell is pushed towards the northwest by this wind.

It would appear that further studies in the few micron wavelength range could
add considerably to the knowledge of the properties of the stellar population in
the Galactic Center.
This is especially true for the $M$-band, since it is a fairly long wavelength
that is not completely dominated by dust emission. The stars therefore remain
accessible for observation.

\begin{acknowledgements}

This work was supported in part by the Deutsche Forschungsgemeinschaft
(DFG) via grant SFB 494.
We are grateful to all members of the NAOS/CONICA
team from MPIA/MPE, Meudon/Grenoble Observatories, ONERA, ESO.
In particular, we thank N.~Ageorges, K.~Bickert, W.~Brandner,
Y.~Cl\'{e}net, E.~Gendron, M.~Hartung, N.~Hubin, C.~Lidman, A.-M.~Lagrange,
A.F.M.~Moorwood, C.~R\"{o}hrle, G.~Rousset and J.~Spyromilio.

\end{acknowledgements}


\bibliographystyle{aa}
\bibliography{ms_Viehmann}


\newpage

\Online
\begin{longtable}{r | l | c c | c c c c | c c c | l}
\caption{Photometry of sources detected in NAOS/CONICA and at least one ISAAC
image. The classifications are from \protect\citet{blum1996}, with ``cool'' referring to red
giants or supergiants and ``red'' signifying objects with featureless $K$-band spectra and a
strong IR excess. The zeropoint for the positions is 17:45:42.93 RA -29:00:29.91
DEC (IRS~16NE) in the J2000 coordinate system, with an offset of 2.83\arcsec
(RA) and $-$0.91\arcsec (DEC) from Sgr~A*.}\\
ID & Name & $\Delta\alpha$(") & $\Delta\delta$(") & $H$ & $K$ & $L$ & $M$ &
$(H-K)_{\mathrm{obs}}$ & $(K-L)_{\mathrm{obs}}$ & $(L-M)_{\mathrm{obs}}$ &
Notes\\
\hline\hline\endfirsthead
\caption{continued}\\
ID & Name & $\Delta\alpha$(") & $\Delta\delta$(") & $H$ & $K$ & $L$ & $M$ &
$(H-K)_{\mathrm{obs}}$ & $(K-L)_{\mathrm{obs}}$ & $(L-M)_{\mathrm{obs}}$ &
Notes\\
\hline\hline\endhead
\hline\endfoot
  50 & IRS~30E & -8.59 & 4.81 & 13.12 & 10.33 & 8.84 & 9.12 & 2.80 & 1.48 & -0.28 & \\
 102 & IRS~9 & 2.67 & -7.51 & 11.78 & 9.41 & 6.89 & 6.90 & 2.38 & 2.52 & -0.02 & cool\\
 179 & & -0.56 & -1.25 & 14.54 & 12.40 & 9.08 & 8.09 & 2.14 & 3.32 & 0.99 & \\
 181 & IRS~30W & -9.40 & 5.17 & 12.86 & 10.21 & 8.76 & 8.87 & 2.65 & 1.45 & -0.12 & \\
 187 & IRS~29 & -4.51 & 0.50 & 14.16 & 10.36 & 6.90 & 6.31 & 3.81 & 3.45 & 0.59 & WC9\\
 197 & & 1.69 & -4.53 & 13.15 & 11.30 & 10.17 & 9.99 & 1.85 & 1.13 & 0.18 & \\
 200 & & -7.53 & -9.83 & 14.46 & 12.09 & 10.65 & $>$12.73 & 2.37 & 1.44 & -3.08 & \\
 202 & IRS~14SW & -3.33 & -10.02 & 11.83 & 9.56 & 8.23 & 8.54 & 2.27 & 1.33 & -0.31 & cool\\
 277 & IRS~16C & -1.79 & -0.49 & 11.70 & 9.64 & 8.07 & 7.78 & 2.06 & 1.57 & 0.29 & He~I\\
 373 & & -11.42 & 6.48 & 13.02 & $>$17.66 & 8.97 & 9.24 & -5.64 & 9.69 & -0.27 & \\
 376 & & -1.45 & -2.44 & 13.76 & 10.84 & 8.92 & 8.23 & 2.92 & 1.92 & 0.69 & \\
 382 & & 0.04 & -6.83 & $>$20.34 & 14.21 & 8.35 & 7.70 & 7.13 & 5.86 & 0.65 & \\
 383 & IRS~12N & -6.38 & -7.87 & 11.73 & 9.25 & 6.63 & 6.43 & 2.48 & 2.63 & 0.19 & cool\\
 385 & IRS~14S & -2.37 & -10.44 & 13.02 & 10.90 & 9.71 & 9.80 & 2.12 & 1.18 & -0.08 & \\
 417 & & -6.82 & 4.09 & 14.25 & 11.73 & 10.08 & 9.37 & 2.52 & 1.65 & 0.72 & \\
 418 & IRS~16CC & -0.87 & -0.49 & 12.13 & 10.19 & 8.50 & 7.92 & 1.94 & 1.69 & 0.57 & He~I\\
 421 & & 0.28 & 5.28 & 14.66 & 11.99 & 10.65 & $>$12.73 & 2.66 & 1.34 & -3.08 & \\
 422 & & 3.23 & 5.90 & 15.06 & 12.59 & 10.89 & $>$12.73 & 2.47 & 1.70 & -2.84 & \\
 423 & IRS~10E & 5.87 & 2.97 & 12.63 & 10.50 & 8.84 & 8.92 & 2.13 & 1.65 & -0.08 & cool\\
 426 & & -10.01 & 2.82 & 14.28 & 12.02 & 11.39 & 9.93 & 2.26 & 0.63 & 1.46 & \\
 427 & & 0.70 & -3.38 & 13.41 & 10.90 & 9.58 & $>$12.73 & 2.50 & 1.32 & -4.15 & \\
 428 & IRS~16NW & -2.86 & 0.26 & 11.66 & 9.84 & 8.38 & 8.24 & 1.82 & 1.46 & 0.14 & He~I\\
 430 & IRS~29S & -4.82 & 0.01 & 12.33 & 10.34 & 9.81 & $>$12.73 & 2.00 & 0.53 & -3.92 & cool\\
 432 & & -11.72 & -2.36 & 14.26 & 11.85 & 10.30 & $>$12.73 & 2.41 & 1.55 & -3.43 & \\
 433 & & -10.34 & -2.53 & 14.73 & 12.34 & 9.38 & 8.43 & 2.39 & 2.96 & 0.95 & \\
 435 & & -10.72 & -4.41 & 13.27 & 11.29 & 9.66 & 9.23 & 1.98 & 1.63 & 0.43 & \\
 437 & & 5.08 & -6.75 & 14.66 & 12.41 & 10.85 & $>$12.73 & 2.25 & 1.56 & -2.88 & \\
 438 & & -9.96 & -8.32 & 13.89 & 11.81 & 10.78 & $>$12.73 & 2.07 & 1.04 & -2.95 & \\
 439 & & -10.25 & -8.80 & 13.09 & 11.50 & 10.28 & 10.11 & 1.59 & 1.22 & 0.17 & \\
 440 & & -10.66 & -8.35 & 14.01 & 11.93 & 10.81 & $>$12.73 & 2.08 & 1.12 & -2.92 & \\
 441 & IRS~14NE & -2.20 & -9.19 & 11.84 & 9.43 & 7.71 & 7.87 & 2.41 & 1.72 & -0.16 & cool\\
 461 & & -4.20 & 1.85 & 14.02 & 11.44 & 9.47 & 9.77 & 2.58 & 1.97 & -0.30 & \\
 462 & & -0.83 & -4.95 & 15.90 & 13.66 & 9.48 & 10.27 & 2.23 & 4.18 & -0.79 & \\
 464 & & 3.89 & -7.63 & 15.32 & 13.14 & 10.85 & $>$12.73 & 2.18 & 2.29 & -2.88 & \\
 465 & & -8.47 & -10.46 & $>$12.73 & 11.63 & 10.43 & 10.85 & 2.11 & 1.20 & -0.42 & \\
 488 & IRS~1C & 4.05 & -0.22 & 12.62 & 10.42 & 8.50 & 8.54 & 2.20 & 1.92 & -0.04 & \\
 490 & & -9.24 & 3.69 & 14.48 & 12.03 & 10.58 & 10.59 & 2.45 & 1.45 & -0.02 & \\
 491 & & -2.37 & -1.44 & 15.16 & 12.88 & 10.38 & $>$12.73 & 2.28 & 2.50 & -3.35 & \\
 492 & IRS~21 & -0.58 & -3.80 & 14.01 & 10.41 & 6.29 & 5.47 & 3.61 & 4.12 & 0.82 & red\\
 498 & & -8.91 & -4.03 & 14.16 & 11.43 & 10.05 & 9.56 & 2.73 & 1.38 & 0.49 & \\
 500 & & -9.48 & -8.30 & 14.00 & 12.09 & 11.33 & 11.34 & 1.91 & 0.77 & -0.02 & \\
 501 & & -3.21 & -8.25 & 14.20 & 12.17 & 10.64 & $>$12.73 & 2.03 & 1.53 & -3.09 & \\
 502 & & -3.59 & -8.36 & 13.71 & 11.26 & 10.21 & 10.98 & 2.45 & 1.05 & -0.77 & \\
 503 & & -5.05 & -10.21 & 12.93 & 10.65 & 9.03 & 9.32 & 2.28 & 1.62 & -0.29 & \\
 523 & & -6.76 & 0.58 & $>$20.34 & 15.07 & 9.66 & 8.80 & 6.27 & 5.41 & 0.87 & \\
 524 & & -6.08 & 0.67 & 15.09 & 12.93 & 9.80 & $>$12.73 & 2.17 & 3.13 & -3.93 & \\
 525 & IRS~29NE & -3.90 & 1.15 & 14.61 & 11.57 & 8.91 & 8.53 & 3.04 & 2.66 & 0.38 & \\
 527 & & -11.21 & -3.72 & 14.33 & 12.03 & 10.74 & 11.35 & 2.31 & 1.29 & -0.61 & \\
 528 & & -11.96 & -4.24 & 13.71 & 11.91 & 10.51 & 10.34 & 1.80 & 1.40 & 0.17 & \\
 529 & & 1.85 & -6.73 & 13.80 & 11.80 & 9.92 & $>$12.73 & 2.00 & 1.89 & -3.81 & \\
 530 & & 2.50 & -7.03 & 13.62 & 11.65 & 8.34 & 7.51 & 1.96 & 3.31 & 0.84 & \\
 531 & & 1.71 & -9.29 & 14.96 & 12.86 & 10.07 & $>$12.73 & 2.10 & 2.79 & -3.66 & \\
 532 & & -11.74 & -6.16 & 14.52 & 12.55 & 11.03 & 10.45 & 1.96 & 1.52 & 0.58 & \\
 533 & & -12.10 & -6.58 & 12.37 & 10.57 & 9.25 & 9.05 & 1.80 & 1.32 & 0.20 & \\
 534 & IRS~12S & -6.57 & -9.54 & 12.07 & 10.03 & 8.31 & 8.91 & 2.04 & 1.73 & -0.61 & cool\\
 535 & & 2.21 & -10.53 & 14.97 & 12.83 & 10.64 & $>$12.73 & 2.14 & 2.19 & -3.09 & \\
 549 & & -11.94 & -10.03 & 14.76 & 12.77 & 10.35 & 10.56 & 1.98 & 2.42 & -0.21 & \\
 567 & & -5.29 & 5.10 & 15.24 & 12.90 & 11.31 & 10.02 & 2.34 & 1.58 & 1.29 & \\
 568 & & 1.44 & 2.20 & 14.93 & 12.99 & 10.06 & $>$12.73 & 1.94 & 2.93 & -3.67 & \\
 569 & IRS~35 & 0.07 & -2.23 & 14.55 & 11.80 & 8.63 & 9.00 & 2.75 & 3.17 & -0.37 & \\
 570 & & 4.06 & -1.31 & 12.95 & 10.45 & 8.53 & 8.91 & 2.49 & 1.93 & -0.39 & \\
 571 & IRS~1SE & 4.65 & -1.59 & 12.47 & 10.48 & 8.78 & $>$12.73 & 1.99 & 1.71 & -4.95 & cool\\
 574 & & 5.25 & 4.63 & 15.35 & 13.20 & 10.96 & $>$12.73 & 2.14 & 2.24 & -2.76 & \\
 580 & & 6.14 & -0.71 & $>$20.34 & 12.65 & 9.95 & 9.14 & 8.70 & 2.70 & 0.81 & \\
 581 & & 6.63 & -0.89 & $>$20.34 & 12.47 & 10.62 & 9.86 & 8.87 & 1.85 & 0.76 & \\
 582 & & -2.86 & -4.37 & 15.98 & 13.71 & 9.85 & $>$12.73 & 2.27 & 3.86 & -3.88 & \\
 584 & & 2.62 & -9.35 & 13.81 & 11.36 & 9.40 & 8.62 & 2.45 & 1.96 & 0.77 & \\
 588 & & -6.64 & -8.51 & 12.58 & 10.61 & 9.04 & 9.04 & 1.97 & 1.58 & -0.01 & \\
 610 & & -11.49 & -7.01 & 13.39 & 11.67 & 10.49 & 10.18 & 1.72 & 1.18 & 0.31 & \\
 612 & & -5.35 & -9.60 & 13.78 & 11.74 & 10.93 & 11.10 & 2.04 & 0.80 & -0.16 & \\
 617 & & -2.93 & -10.87 & 15.79 & 12.80 & 10.51 & 11.04 & 2.99 & 2.29 & -0.53 & \\
 626 & & 2.24 & 3.59 & $>$20.34 & 13.05 & 9.05 & 8.16 & 8.30 & 4.00 & 0.89 & \\
 628 & IRS~1NEE & 4.87 & 0.45 & 12.76 & 10.56 & 9.23 & 9.57 & 2.19 & 1.34 & -0.34 & \\
 629 & & -6.61 & 0.97 & 14.28 & 11.81 & 8.87 & 8.32 & 2.48 & 2.94 & 0.55 & \\
 630 & IRS~34 & -7.01 & 0.77 & 14.64 & 11.32 & 8.23 & 7.62 & 3.32 & 3.09 & 0.61 & He~I\\
 631 & IRS~33SE & -2.33 & -4.13 & 11.97 & 9.95 & 8.17 & 8.03 & 2.02 & 1.78 & 0.13 & \\
 633 & & -2.23 & -5.09 & 13.87 & 11.94 & 10.38 & $>$12.73 & 1.92 & 1.57 & -3.35 & \\
 635 & & 1.07 & -5.84 & 14.16 & 11.81 & 10.20 & $>$12.73 & 2.35 & 1.61 & -3.53 & \\
 636 & & -1.22 & -7.39 & 12.90 & 10.79 & 9.63 & 10.25 & 2.11 & 1.16 & -0.62 & \\
 637 & AF/AHH & -9.68 & -7.81 & 12.46 & 10.43 & 8.70 & 8.10 & 2.03 & 1.72 & 0.60 & He~I\\
 638 & & -0.50 & -8.80 & 14.37 & 11.90 & 10.53 & $>$12.73 & 2.46 & 1.38 & -3.20 & \\
 647 & IRS~7 & -2.80 & 4.63 & 10.32 & 8.77 & 4.48 & 4.30 & 1.56 & 4.28 & 0.18 & cool M2 I\\
 650 & IRS~6E & -8.16 & -0.07 & 12.57 & 9.58 & 6.34 & 5.71 & 3.00 & 3.24 & 0.62 & WC9\\
 651 & IRS~6W & -10.75 & 0.61 & 12.33 & 10.19 & 7.83 & 6.84 & 2.14 & 2.36 & 0.98 & \\
 655 & IRS~1NE & 4.22 & 0.73 & 12.59 & 10.28 & 8.52 & 8.72 & 2.31 & 1.76 & -0.20 & cool\\
 658 & & 4.65 & 4.99 & 16.53 & 14.10 & 10.71 & $>$12.73 & 2.42 & 3.39 & -3.02 & \\
 661 & & -1.25 & -1.62 & 14.93 & 12.83 & 8.86 & 8.42 & 2.10 & 3.96 & 0.45 & \\
 662 & IRS~16SWE & -1.08 & -2.17 & 13.45 & 10.67 & 7.70 & 7.12 & 2.78 & 2.97 & 0.57 & \\
 663 & & -1.51 & -8.26 & 14.02 & 12.03 & 10.43 & $>$12.73 & 1.99 & 1.60 & -3.30 & \\
 669 & & 4.67 & 1.73 & 15.63 & 13.41 & 11.31 & 9.87 & 2.21 & 2.10 & 1.45 & \\
 670 & & -2.88 & -1.06 & 15.56 & 12.93 & 10.62 & $>$12.73 & 2.63 & 2.31 & -3.11 & \\
 671 & & -1.69 & -2.92 & 12.52 & 10.34 & 8.54 & 7.94 & 2.18 & 1.80 & 0.60 & \\
 672 & & -1.33 & -3.91 & 14.08 & 11.84 & 9.54 & $>$12.73 & 2.24 & 2.30 & -4.19 & \\
 674 & & 0.28 & -3.77 & 15.00 & 12.43 & 10.72 & $>$12.73 & 2.56 & 1.71 & -3.01 & \\
 675 & & 0.17 & -5.96 & 14.22 & 12.09 & 8.64 & 8.37 & 2.13 & 3.45 & 0.27 & \\
 676 & & 1.31 & -8.36 & 16.17 & 13.90 & 9.10 & 7.91 & 2.27 & 4.80 & 1.19 & \\
 677 & & -4.83 & -7.43 & 13.61 & 11.36 & 9.72 & 10.05 & 2.24 & 1.64 & -0.33 & \\
 679 & & 2.91 & 5.08 & 17.54 & 13.92 & 10.39 & $>$12.73 & 3.61 & 3.53 & -3.34 & \\
 680 & & 5.43 & 3.25 & 13.48 & 11.29 & 9.72 & $>$12.73 & 2.19 & 1.57 & -4.01 & \\
 681 & IRS~1SW & 3.50 & -0.84 & 12.73 & 10.43 & 7.33 & 6.54 & 2.30 & 3.10 & 0.79 & \\
 690 & & -3.38 & -3.06 & 17.28 & 14.68 & 8.95 & 8.54 & 2.61 & 5.73 & 0.40 & \\
 691 & & 0.68 & -7.04 & 17.28 & $>$17.66 & 9.20 & 8.40 & -1.37 & 9.46 & 0.81 & \\
 694 & & -2.87 & 3.34 & 16.79 & 14.05 & 9.39 & $>$12.73 & 2.74 & 4.66 & -4.34 & \\
 697 & IRS~1W & 2.43 & -0.50 & 11.73 & 8.90 & 4.92 & 4.02 & 2.83 & 3.98 & 0.90 & red\\
 698 & IRS~16NE & 0.00 & 0.00 & 10.68 & 9.18 & 7.16 & 6.87 & 1.49 & 2.03 & 0.29 & He~I\\
 702 & IRS~10W & 3.75 & 4.07 & 12.95 & 9.86 & 6.25 & 5.02 & 3.09 & 3.61 & 1.23 & \\
 704 & & 1.60 & 4.26 & 13.60 & 11.34 & 9.11 & $>$12.73 & 2.26 & 2.23 & -4.62 & \\
 705 & & -9.65 & 2.04 & 17.36 & 14.52 & 9.75 & $>$12.73 & 2.84 & 4.77 & -3.98 & \\
 710 & & -0.06 & -1.12 & 14.77 & 12.34 & 9.11 & $>$12.73 & 2.42 & 3.23 & -4.62 & \\
 711 & IRS~33SW & -3.35 & -4.20 & 12.73 & 10.69 & 8.61 & 8.65 & 2.05 & 2.08 & -0.04 & \\
 715 & & -2.75 & 5.53 & 15.77 & 14.50 & 10.66 & $>$12.73 & 1.27 & 3.84 & -3.07 & \\
 716 & & -0.27 & -5.51 & 17.44 & $>$17.66 & 10.05 & 8.66 & -1.22 & 8.61 & 1.39 & \\
 718 & & 0.83 & -8.37 & 13.85 & 11.69 & 9.15 & 8.40 & 2.16 & 2.53 & 0.75 & \\
 720 & & -7.01 & -6.46 & 14.38 & 11.60 & 7.63 & 6.54 & 2.78 & 3.97 & 1.10 & \\
 721 & & 0.70 & -6.36 & 17.47 & 15.65 & 9.24 & $>$12.73 & 1.82 & 6.41 & -4.49 & \\
 723 & IRS~7SE & -0.61 & 3.52 & 13.38 & 11.20 & 10.32 & 10.22 & 2.18 & 0.88 & 0.10 & \\
 724 & & 2.94 & 1.13 & 14.29 & 12.07 & 9.91 & $>$12.73 & 2.22 & 2.16 & -3.82 & \\
 725 & & 2.98 & -1.05 & 13.77 & 11.16 & 7.34 & 6.91 & 2.60 & 3.82 & 0.43 & \\
 728 & & -9.00 & -1.92 & 16.49 & 14.16 & 9.80 & 8.94 & 2.32 & 4.36 & 0.86 & \\
 729 & & -10.41 & -0.68 & 15.45 & 13.27 & 10.12 & $>$12.73 & 2.18 & 3.15 & -3.61 & \\
 730 & & 0.84 & 5.61 & 15.56 & 13.64 & 11.10 & 9.77 & 1.92 & 2.54 & 1.33 & \\
 731 & & -4.56 & -1.45 & 14.23 & 12.32 & 9.94 & 9.22 & 1.91 & 2.38 & 0.72 & \\
 733 & & -2.25 & -2.60 & 13.03 & 11.10 & 8.56 & $>$12.73 & 1.94 & 2.54 & -5.17 & \\
 737 & & -1.41 & 3.69 & 14.42 & 11.55 & 9.87 & 10.44 & 2.87 & 1.68 & -0.57 & \\
 739 & & -3.85 & 6.78 & 13.08 & 11.37 & 9.65 & 9.50 & 1.71 & 1.72 & 0.15 & \\
 741 & & -3.81 & -2.53 & 13.46 & 11.24 & 9.25 & 9.03 & 2.23 & 1.99 & 0.21 & \\
 742 & & -6.70 & -4.24 & 18.79 & 15.40 & 8.00 & $>$12.73 & 3.39 & 7.41 & -5.73 & \\
 743 & & -3.25 & -7.10 & 13.10 & 10.80 & 9.26 & 9.10 & 2.29 & 1.54 & 0.16 & \\
 744 & & 2.21 & -8.21 & 13.61 & 11.64 & 9.92 & 9.45 & 1.96 & 1.72 & 0.48 & \\
 745 & & 1.00 & 6.24 & $>$20.34 & 11.02 & 8.91 & 7.94 & 10.33 & 2.10 & 0.97 & \\
 748 & & 4.64 & 3.92 & 16.00 & 13.50 & 10.07 & $>$12.73 & 2.50 & 3.43 & -3.66 & \\
 751 & & -7.07 & -2.98 & 13.68 & 10.89 & 7.97 & 8.00 & 2.78 & 2.92 & -0.03 & \\
 752 & & 4.34 & 5.72 & 15.22 & 13.39 & 10.90 & $>$12.73 & 1.83 & 2.49 & -2.83 & \\
 753 & & -1.87 & 4.89 & 13.74 & 11.17 & 9.48 & 10.02 & 2.56 & 1.69 & -0.54 & \\
 754 & & 1.64 & 0.69 & 16.12 & 14.20 & 9.85 & $>$12.73 & 1.92 & 4.34 & -3.87 & \\
 755 & & -4.57 & -3.82 & 13.52 & 11.57 & 9.32 & 9.66 & 1.95 & 2.25 & -0.34 & \\
 757 & & 1.22 & -9.13 & 14.82 & 12.40 & 9.70 & $>$12.73 & 2.42 & 2.70 & -4.03 & \\
 759 & & -5.80 & 4.61 & 14.80 & 12.41 & 10.48 & $>$12.73 & 2.39 & 1.93 & -3.25 & \\
 760 & & -3.83 & -1.96 & 15.76 & 13.42 & 10.77 & $>$12.73 & 2.34 & 2.64 & -2.95 & \\
 761 & & 0.49 & -1.57 & 15.26 & 12.97 & 9.62 & $>$12.73 & 2.29 & 3.36 & -4.11 & \\
 763 & & 1.13 & -7.82 & 15.08 & 13.23 & 9.33 & $>$12.73 & 1.86 & 3.90 & -4.40 & \\
 769 & & -6.06 & -1.82 & 16.69 & 13.61 & 7.36 & 6.55 & 3.08 & 6.24 & 0.82 & \\
 772 & & 0.17 & 2.45 & 15.37 & 12.26 & 9.77 & $>$12.73 & 3.11 & 2.49 & -3.96 & \\
 774 & & -4.29 & -1.21 & 14.07 & 12.19 & 10.23 & 9.80 & 1.88 & 1.96 & 0.42 & \\
 775 & & -6.35 & -3.78 & 18.20 & 15.06 & 7.57 & 7.19 & 3.15 & 7.49 & 0.38 & \\
 776 & & -7.51 & -7.51 & 17.26 & 15.08 & 9.81 & $>$12.73 & 2.17 & 5.27 & -3.92 & \\
 778 & & -1.25 & 5.56 & 15.84 & 12.87 & 11.03 & 11.16 & 2.96 & 1.85 & -0.13 & \\
 779 & & -5.23 & 1.94 & 18.21 & 14.82 & 8.95 & $>$12.73 & 3.39 & 5.87 & -4.77 & \\
 781 & IRS~2S & -6.75 & -5.61 & 12.83 & 10.18 & 7.92 & 7.53 & 2.65 & 2.26 & 0.38 & cool\\
 782 & IRS~20 & -3.92 & -6.23 & 12.54 & 10.57 & 8.70 & 8.61 & 1.97 & 1.87 & 0.09 & cool\\
 784 & & 2.60 & 1.98 & 13.72 & 11.68 & 9.79 & 9.53 & 2.04 & 1.89 & 0.27 & \\
 785 & & -8.54 & 1.74 & 19.78 & $>$17.66 & 9.57 & 8.25 & 1.12 & 9.09 & 1.32 & \\
 786 & IRS~16SW & -1.87 & -1.97 & 11.53 & 9.80 & 7.92 & 7.53 & 1.74 & 1.88 & 0.39 & He~I\\
 791 & & -6.30 & 3.04 & 17.92 & 16.14 & 9.28 & $>$12.73 & 1.77 & 6.86 & -4.44 & \\
 792 & & -8.95 & -1.41 & 17.42 & 14.36 & 10.06 & 8.96 & 3.06 & 4.30 & 1.11 & \\
 820 & IRS~3 & -5.24 & 2.91 & 14.98 & 10.64 & 4.84 & 3.35 & 4.35 & 5.8 & 1.49 & red\\
 825 & IRS~10E* & 4.94 & 3.11 & 13.62 & 9.76 & 5.95 & 5.20 & 3.86 & 3.81 & 0.75 & \\
 828 & & 1.21 & 1.48 & 17.09 & 14.82 & 10.63 & $>$12.73 & 2.26 & 4.20 & -3.10 & \\
 829 & & 0.88 & 0.60 & 14.54 & 12.10 & 9.20 & 8.31 & 2.44 & 2.90 & 0.89 & \\
 830 & & 0.76 & -0.01 & 15.29 & 12.81 & 9.56 & 8.58 & 2.48 & 3.26 & 0.98 & \\
 831 & & 1.02 & -0.79 & 17.14 & 14.91 & 9.78 & $>$12.73 & 2.22 & 5.13 & -3.94 & \\
 833 & & -9.26 & 0.70 & 18.19 & 15.86 & 9.61 & $>$12.73 & 2.33 & 6.25 & -4.12 & \\
 836 & & -8.89 & -0.81 & 16.75 & 14.26 & 9.88 & $>$12.73 & 2.50 & 4.38 & -3.85 & \\
 838 & & -5.62 & -2.32 & $>$20.34 & 13.26 & 8.16 & 7.07 & 8.09 & 5.09 & 1.09 & \\
 839 & IRS~13E & -6.18 & -2.49 & 12.16 & 9.02 & 6.18 & 5.73 & 3.13 & 2.84 & 0.45 & He~I\\
 840 & & -5.93 & -2.63 & 11.90 & $>$17.66 & 7.19 & 6.63 & -6.76 & 11.47 & 0.56 & \\
 841 & & -7.31 & -2.48 & 14.21 & 12.26 & 8.86 & 8.73 & 1.95 & 3.41 & 0.12 & \\
 845 & & -5.69 & -4.65 & 16.57 & 14.01 & 8.82 & $>$12.73 & 2.56 & 5.19 & -4.91 & \\
 846 & IRS~2L & -6.47 & -4.82 & 14.26 & 10.60 & 6.49 & 5.55 & 3.67 & 4.11 & 0.94 & \\
 848 & & -4.98 & -5.19 & 16.53 & 14.32 & 9.33 & 8.60 & 2.21 & 4.99 & 0.73 & \\
 849 & & -5.54 & -5.32 & 16.83 & 14.61 & 8.77 & 7.80 & 2.21 & 5.84 & 0.97 & \\
 851 & & -4.20 & -5.23 & $>$20.34 & 16.04 & 9.66 & 9.24 & 5.31 & 6.38 & 0.42 & \\
 852 & & -4.76 & -5.87 & 15.58 & 13.47 & 9.15 & $>$12.73 & 2.11 & 4.33 & -4.58 & \\
 856 & & -2.93 & -5.84 & 15.69 & 13.63 & 10.34 & $>$12.73 & 2.06 & 3.29 & -3.39 & \\
 857 & & -3.57 & -6.47 & 14.37 & 12.33 & 9.90 & $>$12.73 & 2.04 & 2.43 & -3.83 & \\
 864 & & 1.69 & 3.96 & 15.03 & 12.39 & $>$14.47 & 8.69 & 2.64 & -3.08 & 6.77 & \\
 865 & & 1.57 & -0.87 & 14.83 & 13.00 & $>$14.47 & 8.92 & 1.83 & -2.47 & 6.54 & \\
 866 & & 0.28 & -1.07 & 14.91 & 12.48 & $>$14.47 & 9.09 & 2.42 & -2.98 & 6.38 & \\
 867 & & -11.30 & -0.91 & 16.28 & 14.12 & $>$14.47 & 9.06 & 2.17 & -1.35 & 6.41 & \\
 868 & & -0.90 & -1.61 & 13.81 & 11.48 & $>$14.47 & 8.62 & 2.33 & -3.99 & 6.85 & \\
 869 & & -5.81 & -3.74 & 14.60 & 12.11 & $>$14.47 & 7.73 & 2.48 & -3.36 & 7.74 & \\
 870 & & -4.51 & -5.76 & 14.53 & 12.27 & $>$14.47 & 8.88 & 2.26 & -3.19 & 6.59 & \\
 871 & & 1.49 & -7.91 & 17.80 & 15.11 & $>$14.47 & 8.84 & 2.68 & -0.36 & 6.62 & \\
 874 & & -8.65 & -0.90 & 17.28 & 14.73 & $>$14.47 & 9.68 & 2.55 & -0.74 & 5.79 & \\
 876 & & -5.66 & 1.93 & 16.15 & 13.94 & $>$14.47 & 8.96 & 2.22 & -1.53 & 6.51 & \\
 877 & & 2.51 & -1.02 & 15.05 & 12.36 & $>$14.47 & 7.54 & 2.70 & -3.11 & 7.93 & \\
 878 & & -7.11 & -3.55 & 16.93 & 15.32 & $>$14.47 & 8.36 & 1.62 & -0.15 & 7.11 & \\
 879 & & 1.10 & -6.90 & 15.70 & 13.78 & $>$14.47 & 8.77 & 1.92 & -1.69 & 6.70 & \\
 880 & & -7.27 & -8.02 & 15.33 & 13.12 & $>$14.47 & 9.55 & 2.21 & -2.35 & 5.91 & \\
 881 & & -6.63 & 2.72 & 17.23 & 14.93 & $>$14.47 & 8.11 & 2.30 & -0.54 & 7.36 & \\
 882 & & 1.38 & 0.32 & 16.97 & 15.19 & $>$14.47 & 8.84 & 1.78 & -0.28 & 6.63 & \\
 884 & & -8.58 & -2.16 & 15.11 & 12.66 & $>$14.47 & 9.56 & 2.45 & -2.81 & 5.91 & \\
 886 & & -2.59 & -4.35 & 14.40 & 12.21 & $>$14.47 & 9.77 & 2.19 & -3.26 & 5.69 & \\
 887 & & -7.44 & -6.89 & 16.99 & 14.79 & $>$14.47 & 8.63 & 2.20 & -0.68 & 6.84 & \\
 888 & & -6.79 & -9.56 & 13.30 & 11.08 & $>$14.47 & 10.46 & 2.22 & -4.39 & 5.01 & \\
 890 & & -2.62 & -3.51 & 15.78 & 14.40 & $>$14.47 & 8.34 & 1.38 & -1.07 & 7.13 & \\
 892 & & 2.21 & 3.05 & 17.36 & 15.11 & $>$14.47 & 8.44 & 2.25 & -0.36 & 7.02 & \\
 893 & & -3.88 & -1.62 & 16.56 & 14.85 & $>$14.47 & 10.49 & 1.71 & -0.62 & 4.98 &
\label{fat}
\end{longtable}

\clearpage
\begin{longtable}{r | c c | c c | c}
\caption{Photometry of sources outside the NAOS/CONICA field of view}\\
ID & $\Delta\alpha$(") & $\Delta\delta$(") & $L$ & $M$ & $(L-M)_{\mathrm{obs}}$\\ 
\hline\hline\endfirsthead
\caption{continued}\\
ID & $\Delta\alpha$(") & $\Delta\delta$(") & $L$ & $M$ & $(L-M)_{\mathrm{obs}}$\\ 
\hline\hline\endhead
\hline\endfoot
  1 & -0.02 & 35.63 & 9.58 & 9.93 & -0.35\\
  3 & -5.29 & 32.78 & 10.30 & 10.65 & -0.35\\
  4 & -22.54 & 33.76 & 9.28 & 9.51 & -0.23\\
  5 & 24.42 & 32.32 & 9.61 & 10.06 & -0.45\\
  7 & -8.24 & 31.64 & 8.96 & 9.32 & -0.36\\
  8 & 13.29 & 29.11 & 9.71 & 10.06 & -0.35\\
  9 & 17.07 & 13.60 & 10.38 & 10.97 & -0.59\\
  10 & -25.77 & -0.17 & 9.93 & 10.39 & -0.47\\
  11 & 23.14 & -10.29 & 9.93 & 10.25 & -0.31\\
  12 & 2.34 & -11.24 & 9.46 & 9.38 & 0.08\\
  14 & 27.03 & -18.73 & 10.59 & 11.09 & -0.51\\
  15 & -25.28 & -25.10 & 10.64 & 11.01 & -0.37\\
  16 & -22.50 & -27.16 & 7.49 & 7.46 & 0.04\\
  18 & -3.17 & 36.47 & 8.35 & 8.55 & -0.19\\
  19 & 27.77 & 33.51 & 10.72 & 11.20 & -0.48\\
  20 & -0.08 & 34.14 & 9.12 & 9.53 & -0.41\\
  23 & -33.84 & 32.93 & 9.17 & 9.69 & -0.52\\
  24 & -4.42 & 31.34 & 9.52 & 9.54 & -0.01\\
  25 & -22.26 & 30.09 & 8.54 & 8.76 & -0.23\\
  28 & 18.05 & 25.41 & 9.70 & 9.99 & -0.29\\
  29 & -29.77 & 23.60 & 7.66 & 7.93 & -0.27\\
  30 & 14.63 & 21.12 & 10.01 & 10.24 & -0.23\\
  31 & -25.89 & 21.16 & 7.90 & 8.04 & -0.14\\
  32 & -36.30 & 21.40 & 9.15 & 9.77 & -0.62\\
  36 & -26.57 & 19.16 & 9.23 & 9.44 & -0.21\\
  38 & 5.04 & 17.33 & 9.01 & 9.09 & -0.09\\
  39 & -4.22 & 16.46 & 8.53 & 8.59 & -0.07\\
  40 & -19.59 & 17.13 & 10.24 & 10.25 & -0.01\\
  41 & -19.76 & 15.10 & 8.35 & 8.63 & -0.28\\
  42 & -38.36 & 14.80 & 9.38 & 9.79 & -0.41\\
  43 & 3.20 & 13.79 & 8.87 & 9.20 & -0.33\\
  44 & -32.51 & 12.76 & 10.51 & 11.05 & -0.53\\
  45 & -35.02 & 11.05 & 9.27 & 9.76 & -0.49\\
  47 & 10.37 & 9.59 & 8.36 & 7.29 & 1.07\\
  48 & -33.23 & 7.74 & 8.83 & 9.02 & -0.19\\
  49 & 17.42 & 5.13 & 10.62 & 10.97 & -0.35\\
  51 & 17.84 & 3.86 & 9.50 & 9.84 & -0.34\\
  52 & -13.61 & 2.89 & 9.96 & 9.91 & 0.05\\
  53 & 17.89 & 1.93 & 9.14 & 9.47 & -0.33\\
  54 & -33.00 & 1.67 & 9.45 & 9.87 & -0.42\\
  55 & 12.49 & -1.78 & 10.56 & 11.07 & -0.52\\
  57 & 25.23 & -7.03 & 9.66 & 9.95 & -0.29\\
  58 & 14.12 & -7.76 & 10.60 & 11.19 & -0.59\\
  60 & 23.98 & -9.03 & 9.66 & 9.90 & -0.25\\
  61 & -24.53 & -10.02 & 10.33 & 10.65 & -0.32\\
  62 & 16.47 & -11.64 & 9.99 & 10.38 & -0.39\\
  63 & 33.26 & -12.39 & 9.51 & 9.86 & -0.35\\
  64 & 18.41 & -13.38 & 10.07 & 10.52 & -0.44\\
  65 & -22.71 & -13.90 & 10.42 & 10.63 & -0.21\\
  66 & 10.99 & -17.37 & 10.35 & 11.16 & -0.81\\
  67 & 29.62 & -14.17 & 9.51 & 9.85 & -0.34\\
  68 & -12.15 & -15.74 & 9.44 & 9.63 & -0.19\\
  69 & 24.60 & -14.69 & 9.45 & 9.82 & -0.36\\
  70 & 2.68 & -15.54 & 9.68 & 10.33 & -0.65\\
  71 & -19.19 & -15.94 & 10.34 & 10.72 & -0.38\\
  72 & 23.15 & -17.55 & 8.91 & 9.20 & -0.29\\
  73 & -32.50 & -17.29 & 9.89 & 10.22 & -0.33\\
  74 & -4.43 & -20.02 & 9.91 & 10.33 & -0.42\\
  76 & -23.65 & -21.25 & 9.81 & 10.17 & -0.37\\
  77 & 10.95 & -23.19 & 9.53 & 10.04 & -0.51\\
  79 & -3.74 & -26.49 & 9.56 & 9.92 & -0.35\\
  82 & 25.07 & -28.98 & 10.42 & 10.62 & -0.21\\
  84 & -9.49 & -32.13 & 10.27 & 10.48 & -0.21\\
  86 & 5.16 & -33.02 & 10.25 & 10.89 & -0.64\\
  87 & -25.55 & -33.40 & 9.88 & 10.31 & -0.43\\
  88 & 30.26 & -35.49 & 10.39 & 10.96 & -0.57\\
  89 & 4.65 & -34.95 & 9.71 & 10.19 & -0.48\\
  93 & 17.26 & 27.79 & 8.63 & 8.84 & -0.22\\
  94 & -13.43 & 24.89 & 10.51 & 10.88 & -0.37\\
  95 & -13.68 & 24.10 & 9.05 & 9.38 & -0.33\\
  96 & -16.60 & 20.03 & 7.62 & 7.48 & 0.14\\
  97 & 23.22 & 19.41 & 10.62 & 10.81 & -0.18\\
  98 & 8.58 & -1.75 & 9.51 & 9.68 & -0.17\\
  99 & -15.48 & 7.47 & 9.38 & 9.51 & -0.13\\
 100 & -28.06 & -1.08 & 9.77 & 10.14 & -0.37\\
 101 & -21.11 & -2.56 & 9.96 & 10.83 & -0.87\\
 105 & -28.14 & -14.91 & 9.94 & 10.28 & -0.34\\
 106 & -33.20 & -21.81 & 9.50 & 9.90 & -0.40\\
 107 & 26.39 & -23.40 & 9.62 & 10.05 & -0.43\\
 108 & -2.09 & -28.81 & 7.92 & 8.00 & -0.08\\
 110 & 27.84 & -33.49 & 10.06 & 10.71 & -0.65\\
 113 & 0.14 & 36.45 & 10.24 & 10.66 & -0.41\\
 114 & 1.51 & 35.89 & 10.67 & 11.26 & -0.59\\
 116 & 14.81 & 33.75 & 9.76 & 10.11 & -0.35\\
 117 & -5.84 & 34.15 & 9.98 & 10.34 & -0.36\\
 121 & 11.14 & 30.48 & 10.53 & 10.81 & -0.28\\
 122 & 12.12 & 30.22 & 9.71 & 9.74 & -0.02\\
 123 & 33.68 & 29.52 & 6.35 & 6.29 & 0.06\\
 125 & 13.88 & 29.72 & 9.60 & 9.83 & -0.23\\
 126 & 14.54 & 28.98 & 10.69 & 11.21 & -0.51\\
 127 & 1.32 & 26.33 & 10.04 & 10.29 & -0.25\\
 128 & 2.28 & 26.18 & 10.28 & 10.52 & -0.24\\
 129 & -9.55 & 29.10 & 9.34 & 9.67 & -0.34\\
 131 & 23.22 & 24.94 & 10.44 & 10.76 & -0.32\\
 132 & 27.82 & 25.36 & 10.25 & 10.62 & -0.38\\
 133 & 24.44 & 23.30 & 10.45 & 10.58 & -0.13\\
 134 & 25.25 & 22.75 & 10.56 & 10.95 & -0.40\\
 135 & 21.36 & 24.87 & 10.19 & 10.50 & -0.31\\
 136 & -17.57 & 24.49 & 10.00 & 9.91 & 0.09\\
 137 & 12.69 & 24.28 & 9.37 & 9.61 & -0.24\\
 138 & 18.14 & 23.94 & 9.42 & 9.66 & -0.24\\
 140 & 0.44 & 23.35 & 10.78 & 10.68 & 0.10\\
 142 & -11.32 & 23.01 & 8.84 & 8.86 & -0.01\\
 143 & 16.35 & 21.31 & 9.79 & 10.05 & -0.26\\
 144 & 6.11 & 19.47 & 9.58 & 9.91 & -0.32\\
 145 & 6.76 & 21.14 & 10.23 & 10.45 & -0.22\\
 146 & 24.78 & 20.24 & 9.59 & 9.97 & -0.38\\
 148 & -2.50 & 19.22 & 9.93 & 10.25 & -0.32\\
 149 & 14.35 & 18.70 & 9.26 & 9.61 & -0.36\\
 150 & -37.44 & 19.59 & 9.78 & 10.05 & -0.27\\
 151 & -11.88 & 18.48 & 9.75 & 9.70 & 0.05\\
 152 & -33.03 & 18.90 & 8.53 & 8.73 & -0.20\\
 154 & -25.92 & 17.39 & 10.02 & 10.50 & -0.48\\
 155 & -25.73 & 16.71 & 8.49 & 8.74 & -0.25\\
 157 & -13.82 & 16.91 & 9.99 & 10.14 & -0.15\\
 158 & 28.46 & 15.67 & 9.83 & 9.81 & 0.03\\
 161 & 6.71 & 14.37 & 10.10 & 10.29 & -0.19\\
 162 & 9.12 & 14.17 & 10.87 & 11.22 & -0.35\\
 163 & 16.57 & 12.19 & 10.11 & 10.40 & -0.29\\
 164 & 12.43 & 12.29 & 9.34 & 9.67 & -0.33\\
 166 & 9.09 & 11.60 & 10.03 & 10.14 & -0.11\\
 168 & 10.53 & 11.94 & 8.67 & 8.86 & -0.19\\
 170 & -1.46 & 10.46 & 7.09 & 6.78 & 0.31\\
 171 & -12.33 & 11.29 & 7.95 & 8.08 & -0.13\\
 172 & -3.29 & 10.90 & 8.83 & 9.10 & -0.28\\
 174 & -5.04 & 7.64 & 8.59 & 8.68 & -0.09\\
 176 & -8.49 & 8.66 & 8.92 & 9.29 & -0.37\\
 178 & -17.60 & 6.84 & 8.78 & 8.67 & 0.11\\
 180 & -33.05 & 7.00 & 10.22 & 10.65 & -0.42\\
 184 & 13.05 & 3.47 & 10.78 & 11.89 & -1.11\\
 185 & 12.63 & 0.96 & 10.07 & 10.32 & -0.26\\
 186 & -34.85 & 3.25 & 9.45 & 10.02 & -0.57\\
 188 & 23.19 & 0.32 & 9.91 & 10.41 & -0.50\\
 189 & 21.88 & 0.07 & 10.40 & 11.87 & -1.47\\
 190 & 15.32 & -0.42 & 10.65 & 11.92 & -1.27\\
 191 & 20.00 & -1.49 & 8.97 & 9.19 & -0.22\\
 192 & -16.42 & -1.86 & 10.29 & 11.29 & -0.99\\
 193 & -16.53 & -2.64 & 9.96 & 9.93 & 0.04\\
 194 & 12.69 & -5.56 & 9.48 & 9.57 & -0.09\\
 195 & 10.76 & -6.34 & 9.64 & 9.73 & -0.09\\
 201 & -9.35 & -11.85 & 10.77 & 10.48 & 0.30\\
 203 & -33.34 & -10.05 & 9.93 & 10.34 & -0.41\\
 204 & 22.40 & -12.51 & 9.06 & 9.38 & -0.32\\
 205 & -17.24 & -13.39 & 9.94 & 10.16 & -0.22\\
 208 & 0.50 & -12.50 & 7.79 & 8.14 & -0.34\\
 209 & -19.54 & -13.13 & 10.80 & 11.68 & -0.88\\
 210 & 3.31 & -14.02 & 10.79 & 12.02 & -1.22\\
 211 & -0.93 & -16.21 & 8.97 & 9.19 & -0.23\\
 212 & -16.15 & -14.22 & 10.26 & 10.45 & -0.19\\
 213 & -3.11 & -16.20 & 9.63 & 10.02 & -0.39\\
 214 & -32.14 & -16.10 & 9.76 & 10.14 & -0.38\\
 215 & -2.60 & -18.12 & 10.44 & 11.10 & -0.65\\
 219 & 2.33 & -19.32 & 9.45 & 9.78 & -0.33\\
 220 & -4.05 & -20.36 & 10.05 & 10.39 & -0.34\\
 221 & -23.00 & -18.57 & 9.45 & 9.75 & -0.31\\
 222 & 17.12 & -20.15 & 9.62 & 9.99 & -0.36\\
 224 & -15.50 & -19.85 & 9.41 & 9.75 & -0.34\\
 225 & -26.91 & -20.76 & 10.19 & 10.51 & -0.31\\
 226 & -9.41 & -22.53 & 8.48 & 8.51 & -0.03\\
 227 & -9.74 & -22.14 & 9.48 & 9.91 & -0.43\\
 228 & -24.79 & -22.45 & 10.04 & 10.28 & -0.24\\
 231 & 0.66 & -23.28 & 10.17 & 10.24 & -0.07\\
 232 & -5.56 & -23.05 & 10.47 & 10.73 & -0.26\\
 234 & 26.56 & -25.33 & 9.27 & 9.65 & -0.37\\
 235 & -22.85 & -25.24 & 9.31 & 9.63 & -0.32\\
 236 & -1.15 & -26.21 & 10.32 & 10.58 & -0.26\\
 237 & -38.85 & -25.91 & 9.41 & 10.41 & -1.00\\
 238 & -2.33 & -26.55 & 10.07 & 10.44 & -0.37\\
 239 & 0.55 & -27.68 & 10.34 & 10.70 & -0.36\\
 240 & 0.93 & -31.02 & 10.51 & 10.95 & -0.44\\
 241 & -5.92 & -30.80 & 10.44 & 10.88 & -0.43\\
 242 & -30.11 & -30.94 & 8.95 & 9.20 & -0.25\\
 243 & -19.09 & -31.23 & 10.40 & 10.64 & -0.24\\
 244 & -8.56 & -33.04 & 10.27 & 10.55 & -0.28\\
 245 & -30.87 & -33.15 & 9.78 & 10.29 & -0.51\\
 248 & -18.52 & -35.56 & 9.73 & 10.04 & -0.31\\
 250 & 12.07 & 34.82 & 9.35 & 9.81 & -0.47\\
 251 & 32.39 & 33.37 & 7.97 & 8.20 & -0.23\\
 254 & 2.06 & 31.25 & 10.59 & 10.84 & -0.24\\
 257 & 32.79 & 28.90 & 9.55 & 9.86 & -0.31\\
 258 & -13.12 & 25.62 & 8.82 & 9.15 & -0.33\\
 259 & -13.83 & 25.46 & 9.65 & 9.92 & -0.28\\
 262 & -14.90 & 25.05 & 9.88 & 10.27 & -0.40\\
 263 & 11.82 & 23.24 & 9.82 & 10.14 & -0.32\\
 264 & -18.96 & 23.20 & 7.89 & 8.05 & -0.16\\
 267 & -13.35 & 21.90 & 10.75 & 11.74 & -1.00\\
 268 & 19.37 & 19.10 & 10.95 & 11.07 & -0.11\\
 269 & 16.78 & 17.19 & 9.73 & 9.99 & -0.26\\
 272 & -4.72 & 12.96 & 10.24 & 10.60 & -0.36\\
 274 & 19.08 & 11.30 & 10.30 & 10.86 & -0.56\\
 276 & 13.68 & 7.76 & 9.30 & 9.69 & -0.40\\
 278 & -2.16 & 8.33 & 9.45 & 9.93 & -0.48\\
 279 & -2.77 & 7.68 & 9.42 & 9.39 & 0.03\\
 280 & 9.29 & 5.90 & 10.31 & 10.99 & -0.68\\
 281 & -35.04 & 5.27 & 9.43 & 9.94 & -0.51\\
 282 & 10.51 & 4.38 & 6.99 & 7.02 & -0.03\\
 283 & 13.02 & 4.63 & 9.83 & 10.23 & -0.40\\
 284 & 20.56 & 0.22 & 10.29 & 10.62 & -0.33\\
 285 & 7.76 & 0.04 & 8.18 & 8.33 & -0.15\\
 289 & 19.18 & -2.40 & 9.87 & 10.19 & -0.32\\
 291 & 11.54 & -8.24 & 10.22 & 10.17 & 0.05\\
 292 & -13.88 & -8.94 & 9.22 & 9.60 & -0.38\\
 293 & 5.83 & -9.73 & 10.90 & 11.58 & -0.68\\
 294 & 8.36 & -10.23 & 8.96 & 9.08 & -0.12\\
 296 & -24.10 & -11.47 & 10.51 & 10.60 & -0.09\\
 297 & 25.95 & -13.76 & 10.14 & 10.53 & -0.39\\
 298 & 25.70 & -14.13 & 10.48 & 10.88 & -0.40\\
 299 & -22.03 & -13.56 & 10.44 & 10.71 & -0.27\\
 302 & -6.79 & -16.75 & 8.15 & 8.30 & -0.15\\
 304 & -8.96 & -17.43 & 9.21 & 9.17 & 0.04\\
 305 & -9.39 & -18.53 & 9.23 & 9.56 & -0.33\\
 306 & -8.27 & -20.20 & 10.20 & 11.18 & -0.99\\
 307 & 16.01 & -18.66 & 9.60 & 10.05 & -0.45\\
 308 & -11.08 & -20.18 & 10.48 & 10.49 & -0.01\\
 309 & -34.68 & -20.77 & 9.58 & 9.92 & -0.34\\
 310 & 12.08 & -22.50 & 9.37 & 9.91 & -0.54\\
 311 & -10.85 & -24.26 & 10.82 & 11.09 & -0.27\\
 313 & -24.69 & -25.40 & 10.18 & 10.53 & -0.35\\
 314 & -9.35 & -26.24 & 10.48 & 10.78 & -0.30\\
 316 & -8.63 & -27.87 & 10.34 & 10.79 & -0.44\\
 317 & -33.27 & -31.73 & 9.89 & 10.16 & -0.27\\
 318 & -16.83 & -35.78 & 10.26 & 10.30 & -0.04\\
 319 & -30.69 & -35.56 & 7.74 & 7.70 & 0.04\\
 322 & -26.47 & 34.82 & 9.10 & 9.61 & -0.51\\
 323 & 27.85 & 31.55 & 10.82 & 10.67 & 0.15\\
 324 & 17.65 & 27.16 & 10.44 & 10.38 & 0.06\\
 325 & -5.61 & 18.99 & 9.18 & 9.24 & -0.06\\
 329 & 8.85 & 2.61 & 8.75 & 9.01 & -0.27\\
 332 & 14.37 & -6.20 & 8.52 & 8.69 & -0.17\\
 333 & -27.66 & -6.31 & 9.73 & 10.41 & -0.68\\
 334 & 17.07 & -11.21 & 9.98 & 10.20 & -0.22\\
 335 & -15.80 & -11.94 & 7.75 & 7.83 & -0.09\\
 336 & -22.72 & -14.87 & 10.44 & 10.40 & 0.04\\
 339 & -2.01 & -23.02 & 8.30 & 8.37 & -0.07\\
 340 & -27.77 & -26.09 & 10.25 & 10.66 & -0.41\\
 342 & -36.23 & 36.07 & 9.68 & 10.22 & -0.54\\
 343 & 2.39 & 33.38 & 8.21 & 8.43 & -0.22\\
 344 & -20.15 & 33.04 & 8.48 & 8.68 & -0.20\\
 346 & -20.70 & 30.38 & 9.89 & 10.06 & -0.17\\
 350 & 12.13 & 23.84 & 10.31 & 10.55 & -0.25\\
 351 & 20.49 & 22.42 & 7.66 & 7.84 & -0.18\\
 352 & -27.42 & 21.31 & 10.20 & 10.65 & -0.45\\
 353 & -16.25 & 20.53 & 10.89 & 11.31 & -0.42\\
 355 & 3.18 & 19.18 & 10.31 & 10.49 & -0.19\\
 356 & 19.39 & 17.89 & 10.91 & 10.83 & 0.08\\
 358 & -0.73 & 14.74 & 9.65 & 10.04 & -0.39\\
 363 & 6.51 & 7.01 & 8.79 & 7.97 & 0.82\\
 366 & -1.27 & 10.83 & 9.96 & 9.27 & 0.69\\
 368 & -7.11 & 10.48 & 9.20 & 9.43 & -0.24\\
 369 & 17.21 & 9.80 & 10.42 & 10.66 & -0.24\\
 371 & 12.74 & 8.04 & 9.55 & 9.93 & -0.39\\
 372 & -10.39 & 8.84 & 10.64 & 10.44 & 0.20\\
 374 & -12.21 & 6.02 & 10.87 & 11.74 & -0.87\\
 380 & -15.93 & -5.36 & 9.67 & 9.95 & -0.28\\
 381 & -13.13 & -5.51 & 10.70 & 10.93 & -0.23\\
 384 & 8.17 & -7.88 & 9.20 & 9.17 & 0.02\\
 386 & -1.56 & -12.39 & 10.78 & 10.69 & 0.09\\
 387 & 8.28 & -11.39 & 9.87 & 9.74 & 0.13\\
 388 & -12.97 & -12.19 & 9.85 & 10.32 & -0.48\\
 389 & -21.51 & -12.73 & 10.45 & 10.79 & -0.34\\
 390 & -10.92 & -13.99 & 10.33 & 10.91 & -0.58\\
 391 & -11.43 & -14.63 & 9.47 & 9.81 & -0.34\\
 392 & -1.16 & -14.98 & 9.91 & 10.16 & -0.25\\
 393 & -1.82 & -15.53 & 10.72 & 10.91 & -0.19\\
 395 & -6.03 & -18.78 & 10.94 & 11.21 & -0.27\\
 396 & -31.77 & -18.34 & 8.60 & 8.79 & -0.19\\
 397 & -11.38 & -18.95 & 10.88 & 11.97 & -1.08\\
 398 & -12.21 & -20.76 & 8.54 & 8.68 & -0.14\\
 399 & -22.27 & -18.74 & 9.42 & 9.74 & -0.32\\
 400 & 1.17 & -19.78 & 9.73 & 10.14 & -0.41\\
 401 & -13.29 & -29.36 & 9.59 & 9.97 & -0.37\\
 403 & -32.11 & -30.38 & 8.98 & 9.20 & -0.21\\
 404 & 14.24 & 31.98 & 10.64 & 10.86 & -0.23\\
 406 & 0.82 & 25.80 & 9.76 & 10.06 & -0.30\\
 407 & -16.23 & 22.79 & 9.91 & 10.24 & -0.32\\
 408 & -8.23 & 22.46 & 10.85 & 11.59 & -0.73\\
 409 & -0.88 & 21.12 & 9.85 & 10.04 & -0.19\\
 410 & 0.61 & 17.67 & 9.03 & 9.12 & -0.09\\
 411 & -11.36 & 15.26 & 10.47 & 10.09 & 0.38\\
 412 & -11.22 & 12.91 & 7.13 & 7.34 & -0.21\\
 413 & -10.92 & 12.89 & 9.08 & 9.03 & 0.04\\
 415 & -8.84 & 10.68 & 10.75 & 11.37 & -0.62\\
 416 & -13.09 & 8.76 & 9.25 & 9.47 & -0.22\\
 419 & 9.34 & -2.23 & 9.46 & 9.83 & -0.37\\
 420 & -35.35 & 9.20 & 9.64 & 10.12 & -0.48\\
 429 & -31.13 & 2.22 & 10.21 & 11.12 & -0.91\\
 434 & -12.77 & -3.34 & 10.18 & 10.17 & 0.01\\
 436 & 27.93 & -5.47 & 10.71 & 11.23 & -0.52\\
 442 & 14.67 & -9.68 & 9.98 & 10.49 & -0.51\\
 443 & -12.38 & -9.85 & 10.39 & 11.03 & -0.64\\
 444 & -8.53 & -13.54 & 10.71 & 10.65 & 0.06\\
 445 & 10.27 & -14.33 & 9.77 & 10.57 & -0.80\\
 447 & 12.19 & -17.47 & 9.73 & 10.67 & -0.94\\
 448 & 24.69 & -19.48 & 8.87 & 9.07 & -0.20\\
 450 & -7.03 & -25.10 & 9.05 & 9.36 & -0.31\\
 454 & -7.29 & -28.62 & 8.18 & 8.39 & -0.21\\
 455 & -18.30 & -31.53 & 9.60 & 9.85 & -0.25\\
 456 & 18.67 & 19.87 & 8.77 & 8.87 & -0.10\\
 457 & -2.95 & 17.42 & 10.76 & 10.79 & -0.02\\
 458 & 12.95 & 9.32 & 10.31 & 10.75 & -0.45\\
 459 & 1.33 & -0.09 & 9.24 & 8.84 & 0.40\\
 460 & 12.82 & 1.91 & 10.12 & 10.45 & -0.33\\
 463 & -28.82 & -5.54 & 10.28 & 10.76 & -0.47\\
 466 & -7.46 & -23.25 & 8.54 & 8.51 & 0.03\\
 467 & -5.67 & -28.77 & 9.71 & 10.00 & -0.29\\
 469 & 1.55 & 33.58 & 9.37 & 9.68 & -0.30\\
 470 & 2.74 & 33.91 & 10.38 & 10.57 & -0.20\\
 474 & -4.26 & 18.93 & 9.35 & 9.51 & -0.16\\
 475 & -5.13 & 16.62 & 10.42 & 11.00 & -0.59\\
 478 & 5.80 & 11.81 & 9.10 & 9.43 & -0.33\\
 479 & 1.85 & 10.11 & 10.66 & 10.72 & -0.06\\
 483 & 8.14 & 5.71 & 9.48 & 8.91 & 0.57\\
 486 & -2.61 & 9.40 & 9.67 & 9.52 & 0.15\\
 487 & -6.18 & 8.87 & 8.81 & 9.08 & -0.27\\
 489 & 26.64 & 6.98 & 9.85 & 10.33 & -0.48\\
 493 & -22.32 & -1.51 & 10.31 & 10.62 & -0.32\\
 494 & 17.85 & -3.72 & 8.71 & 8.80 & -0.09\\
 495 & 18.79 & -3.40 & 10.52 & 10.90 & -0.39\\
 496 & 18.14 & -4.41 & 9.65 & 10.09 & -0.45\\
 499 & 10.02 & -7.97 & 10.66 & 11.10 & -0.44\\
 506 & -38.26 & -14.45 & 9.04 & 9.40 & -0.36\\
 507 & -12.93 & -16.43 & 9.90 & 9.91 & -0.01\\
 508 & 14.10 & -23.96 & 10.28 & 10.93 & -0.65\\
 509 & -31.21 & -28.27 & 9.28 & 9.58 & -0.30\\
 511 & -10.46 & -31.71 & 10.60 & 10.88 & -0.28\\
 512 & 15.63 & 27.39 & 9.89 & 10.13 & -0.24\\
 513 & 10.75 & 22.49 & 8.37 & 8.62 & -0.25\\
 515 & -4.49 & 12.33 & 10.53 & 11.09 & -0.56\\
 516 & 4.48 & 9.03 & 9.46 & 9.42 & 0.04\\
 517 & -4.30 & 9.21 & 9.84 & 9.22 & 0.61\\
 520 & 23.48 & 2.33 & 8.83 & 9.00 & -0.17\\
 521 & 15.03 & 2.42 & 10.79 & 10.89 & -0.10\\
 526 & 22.70 & -2.19 & 10.73 & 11.17 & -0.43\\
 536 & -9.12 & -10.83 & 9.54 & 9.86 & -0.32\\
 538 & 7.82 & -15.03 & 9.73 & 9.98 & -0.25\\
 539 & 11.44 & -20.03 & 7.77 & 7.94 & -0.17\\
 540 & -17.83 & 27.85 & 9.78 & 9.94 & -0.16\\
 541 & 5.89 & 20.15 & 10.11 & 10.31 & -0.20\\
 542 & 13.67 & 9.17 & 10.33 & 10.72 & -0.39\\
 543 & 13.54 & 8.83 & 9.98 & 10.36 & -0.38\\
 551 & 12.29 & -12.25 & 7.82 & 8.02 & -0.20\\
 552 & -11.89 & -12.12 & 8.85 & 8.95 & -0.10\\
 553 & -36.86 & -13.97 & 8.50 & 8.68 & -0.18\\
 554 & 13.04 & -23.34 & 7.87 & 7.98 & -0.11\\
 555 & -1.71 & -23.66 & 8.91 & 9.08 & -0.18\\
 556 & -6.81 & -25.45 & 9.07 & 9.46 & -0.39\\
 559 & 16.62 & 29.76 & 10.44 & 10.89 & -0.45\\
 560 & 1.70 & 28.06 & 9.32 & 9.53 & -0.21\\
 562 & -15.81 & 19.23 & 10.81 & 10.66 & 0.15\\
 563 & 10.64 & 15.65 & 10.73 & 10.05 & 0.68\\
 564 & -2.17 & 17.05 & 9.44 & 9.27 & 0.17\\
 565 & -13.49 & 13.25 & 10.68 & 10.07 & 0.61\\
 572 & -11.53 & 7.65 & 10.65 & 10.52 & 0.14\\
 576 & -17.99 & 3.44 & 10.39 & 10.47 & -0.08\\
 577 & -18.14 & 0.63 & 9.99 & 10.07 & -0.08\\
 583 & -12.52 & -1.29 & 9.27 & 9.15 & 0.12\\
 586 & 7.56 & -7.09 & 7.30 & 7.43 & -0.13\\
 587 & 32.91 & -10.06 & 9.55 & 9.92 & -0.37\\
 589 & -25.64 & -9.06 & 10.99 & 10.80 & 0.19\\
 592 & 5.25 & -18.40 & 10.44 & 10.49 & -0.05\\
 593 & -37.08 & -18.52 & 8.36 & 8.08 & 0.28\\
 596 & 26.90 & 26.26 & 9.31 & 9.75 & -0.45\\
 598 & -10.05 & 19.58 & 9.71 & 9.58 & 0.12\\
 599 & 4.20 & 12.27 & 10.44 & 10.02 & 0.42\\
 602 & 3.21 & 10.63 & 9.83 & 10.10 & -0.27\\
 606 & 11.97 & 9.27 & 9.74 & 10.04 & -0.30\\
 607 & 0.35 & -0.65 & 8.93 & 8.12 & 0.81\\
 608 & 20.50 & 1.97 & 9.05 & 9.13 & -0.08\\
 614 & 14.67 & 4.26 & 8.94 & 9.43 & -0.49\\
 615 & 23.99 & 1.72 & 10.00 & 10.10 & -0.11\\
 616 & -33.14 & -6.46 & 10.45 & 10.67 & -0.22\\
 618 & -32.77 & -12.34 & 9.60 & 9.93 & -0.33\\
 619 & -34.81 & -12.20 & 9.38 & 9.64 & -0.26\\
 620 & 24.33 & -23.45 & 9.75 & 9.89 & -0.15\\
 622 & -19.58 & 28.37 & 10.22 & 10.41 & -0.19\\
 623 & -7.29 & 24.98 & 10.48 & 10.95 & -0.47\\
 624 & 8.35 & 8.33 & 9.71 & 8.76 & 0.94\\
 625 & 2.32 & 4.12 & 9.93 & 8.58 & 1.35\\
 627 & -15.97 & 3.17 & 10.97 & 11.05 & -0.08\\
 640 & 1.89 & 31.98 & 9.84 & 10.10 & -0.26\\
 643 & -8.30 & 12.85 & 10.32 & 10.69 & -0.37\\
 645 & 4.67 & 11.66 & 10.56 & 9.99 & 0.57\\
 648 & 14.96 & 4.21 & 8.54 & 8.55 & -0.01\\
 653 & 8.86 & -15.74 & 10.55 & 11.13 & -0.59\\
 657 & 20.72 & -19.77 & 8.66 & 9.05 & -0.38\\
 678 & -6.22 & -27.05 & 6.67 & 6.91 & -0.24\\
 684 & 0.79 & -21.58 & 10.59 & 10.76 & -0.17\\
 685 & 17.78 & 19.01 & 9.25 & 9.43 & -0.18\\
 686 & 19.89 & 14.06 & 8.67 & 9.02 & -0.35\\
 687 & -3.28 & 8.90 & 9.97 & 10.39 & -0.42\\
 692 & 18.98 & -11.26 & 9.86 & 9.90 & -0.04\\
 693 & -18.19 & 5.33 & 10.21 & 10.60 & -0.38\\
 699 & -2.50 & 28.93 & 10.12 & 9.29 & 0.83\\
 700 & -2.73 & 28.40 & 10.22 & 9.37 & 0.85\\
 708 & -1.20 & 28.75 & 7.24 & 6.00 & 1.24\\
 709 & 15.45 & 1.47 & 9.76 & 10.10 & -0.34\\
 714 & -5.58 & 10.49 & 8.91 & 9.21 & -0.31\\
 717 & 0.29 & -7.74 & 9.60 & 8.83 & 0.76\\
 727 & -10.32 & 0.82 & 8.56 & 7.04 & 1.52\\
 734 & -3.02 & -3.33 & 8.91 & 8.00 & 0.91\\
 735 & 17.70 & 29.44 & 9.28 & 9.49 & -0.21\\
 747 & -18.45 & 6.09 & 10.26 & 10.33 & -0.07\\
 749 & -11.37 & -29.74 & 7.14 & 7.20 & -0.05\\
 750 & -0.93 & 28.97 & 7.41 & 5.77 & 1.64\\
 756 & -2.71 & -2.96 & 8.84 & 8.16 & 0.68\\
 764 & 6.11 & -19.72 & 7.95 & 8.17 & -0.21\\
 765 & 20.41 & 13.92 & 8.50 & 8.81 & -0.31\\
 767 & -7.94 & 1.55 & 10.64 & 10.11 & 0.54\\
 768 & -8.65 & 0.65 & 9.92 & 8.77 & 1.15\\
 773 & -9.78 & 1.33 & 9.79 & 8.08 & 1.71\\
 793 & 11.72 & -20.26 & 6.66 & 7.04 & -0.38\\
 794 & -1.59 & -31.22 & 8.26 & 8.47 & -0.22\\
 798 & -5.78 & 2.28 & 8.97 & 7.84 & 1.13\\
 800 & -7.13 & -30.04 & 9.93 & 10.19 & -0.26\\
 808 & -26.87 & 23.00 & 8.16 & 8.44 & -0.28\\
 810 & 15.03 & 14.01 & 9.22 & 9.55 & -0.32\\
 811 & 5.97 & 8.80 & 6.10 & 5.33 & 0.78\\
 813 & 25.47 & 8.84 & 8.56 & 8.87 & -0.31\\
 814 & 24.12 & 6.71 & 8.74 & 9.12 & -0.37\\
 819 & -5.28 & 3.06 & 6.54 & 3.35 & 3.19\\
 842 & -5.62 & -3.44 & 8.23 & 7.67 & 0.56\\
 858 & -28.78 & -6.02 & 9.04 & 9.36 & -0.32\\
 859 & -18.53 & -22.65 & 8.97 & 9.24 & -0.27\\
 860 & -13.26 & -27.11 & 8.66 & 8.88 & -0.23\\
 861 & 4.67 & -29.35 & 7.70 & 6.73 & 0.96\\
 862 & -12.01 & -32.07 & 8.45 & 8.88 & -0.43\\
 863 & -39.97 & -33.65 & 10.93 & 9.54 & 1.39
\label{outside}
\end{longtable}

\end{document}